\documentclass[12pt,a4paper, superscriptaddress, onecolumn,a4paper]{revtex4-1}
\usepackage{ulem}
\usepackage{color}
\usepackage{graphicx}
\usepackage{appendix}
\usepackage{epsfig}
\usepackage{epstopdf}
\usepackage{amsmath}
\usepackage{subfig}
\usepackage{comment}
\usepackage{bm}
\usepackage{graphics}
\usepackage{graphicx}
\usepackage{color}
\usepackage[english]{babel}
\usepackage[colorlinks=true,linkcolor= cyan, citecolor = cyan, urlcolor = cyan ]{hyperref}
\newcommand{\be}{\begin{equation}}
\newcommand{\ee}{\end{equation}}
\newcommand{\ba}{\begin{eqnarray}}
\newcommand{\ea}{\end{eqnarray}}
\newcommand{\bi}{\begin{itemize}}
\newcommand{\ei}{\end{itemize}}
\newcommand{\nn}{\nonumber\\}

\begin{document}
\title{Collective excitations in the hot QCD medium and the propagation of Heavy Quarks}

\author{Mohammad Yousuf Jamal}
	\email{yousufjml5@gmail.com}
\affiliation {Key Laboratory of Quark and Lepton Physics (MOE) \& Institute of Particle Physics, Central China Normal University, Wuhan 430079, China}

\author{Bedangadas Mohanty}
	\email{bedanga@niser.ac.in }
\affiliation {School of Physical Sciences, National Institute of Science Education and Research, HBNI, Jatni-752050, India}

\begin{abstract}
This review explores the current understanding of collective excitations and the dynamics of heavy quark propagation in the quark-gluon plasma (QGP) formed in relativistic heavy-ion collisions. We focus on three core aspects: the theoretical modelling of the QGP, including momentum anisotropy, medium-induced collisions, finite chemical potential, and non-ideal interactions; the collective behaviours within the plasma; and the interaction dynamics of heavy quarks as they traverse the medium. Along with the polarization energy loss mechanisms, we also review the possibility of energy gain due to thermal field fluctuations. Lastly, we discuss how these theoretical insights can be tested through experiments and outline possible directions for future research.
\end{abstract}

\maketitle

\tableofcontents
\section{Introduction and motivation}
\label{sec:intro}
\begin{figure}[h]
\centering
\includegraphics[height=7cm,width=7cm]{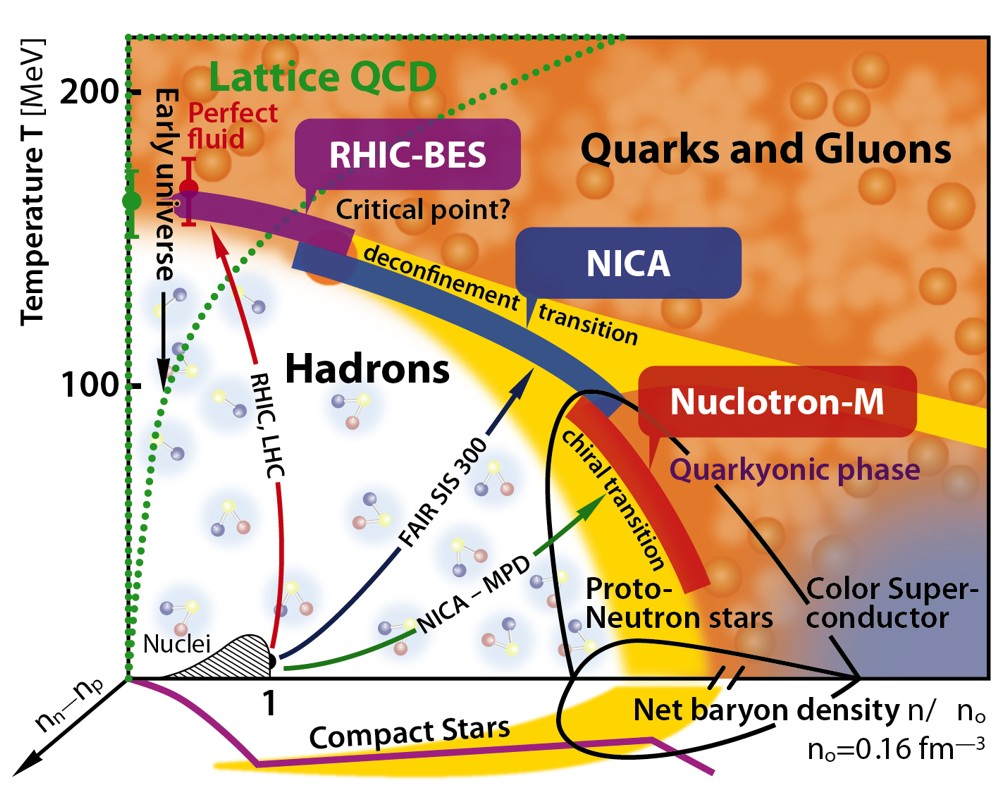}
\caption{Schematic QCD phase diagram for nuclear matter with possible trajectories of created systems and the QGP phase at different accelerator facilities~\cite{Phase}}
\label{fig:Phase}
\end{figure}

The Big Bang theory, established in the 1940s~\cite{ParticleDataGroup:2014cgo}, stands as the predominant framework for understanding the genesis and evolution of the universe. It proposes a progression from a high-density, high-temperature state, where the universe underwent successive phase transitions. The scientific community constructed heavy-ion colliders to scrutinize the Quantum Chromodynamic (QCD) phase transition, facilitating controlled experiments that emulate the primordial Big Bang conditions, often called the ``mini bang"~\cite{STAR:2005gfr, ALICE:2010suc, ALICE:2010mlf}. Fig.~\ref{fig:Phase} illustrates the envisioned phases of the quark and gluon-rich medium based on QCD~\cite{Phase}. The QCD phase diagram, introduced in 1975~\cite{Cabibbo:1975ig}, consists of two major regions: one representing the confining state where quarks and gluons are bound within hadrons and another representing the deconfined state where quarks and gluons move freely beyond hadronic boundaries. A critical temperature, \( T_c \), separates these two states, with experimental investigations suggesting \( T_c \sim 155 \) MeV~\cite{Luo:2011rg, Mohanty:2009vb, STAR:2020tga, STAR:2021iop, STAR:2021rls}. Relativistic heavy-ion collisions (HIC) are crucial for mapping these regions of the QCD phase diagram. A deep understanding of the following sequential stages of evolution of the created matter is required. In HIC, the initial collision of Lorentz-contracted nuclei triggers the QCD vacuum, giving rise to a non-equilibrated medium known as the Glasma~\cite{Gelis:2010nm}. This Glasma state represents the earliest phase of matter formed in these collisions, setting the stage for the subsequent evolution through different phases of QCD matter. During this phase, substantial entropy is generated, and particle density rapidly increases, forming a locally thermally equilibrated state known as the quark-gluon plasma (QGP)~\cite{Shuryak:2004cy}. The QGP exhibits near-perfect fluidic characteristics~\cite{Ryu:2015vwa, Denicol:2015nhu} with a very short lifespan, eventually transmuting into a hadronic state. The nature of the phase transition from the deconfined to confined regions remains under scrutiny, with indications of a simple crossover for smaller chemical potential values~\cite{Borsanyi:2020fev, Tan:2019zyw} and speculations about a first-order transition at low temperatures and high baryon densities~\cite{Scavenius:2000qd, Pradeep:2024cca}. Subsequent phases involve chemical freeze-out, kinetic freeze-out, and finally, a state where spectra of colourless free streaming particles such as hadrons, leptons, and photons are detected. Given the transient nature of the QGP medium, which typically lasts only a few femtometers, it is challenging to use external probes to investigate the produced matter. Consequently, researchers have proposed various self-generated probes to validate QGP production, including phenomena such as strangeness enhancement~\cite{Margetis:2000sv, STAR:2023hwu}, collective flow~\cite{Herrmann:1999wu, Bhatta:2024ygy}, quarkonia dissociation~\cite{Wong:2001uu, Sebastian:2022sga}, and jet quenching~\cite{Qin:2015srf, Shan-Liang:2023sdk}. These probes serve as valuable indicators of QGP formation, providing essential insights into its characteristics.

Due to their significantly higher mass (\(M \gg T\), where \(T\) denotes the temperature of the QGP medium), heavy 
quarks, namely charm and bottom, are not expected to thermalize with the produced medium. Moreover, these quarks are predominantly generated during the initial stages of collisions, with a formation time \(\tau_f \sim 1/M\), making them excellent probes for studying different phases of the created matter. They provide crucial insights into phenomena such as the suppression or enhancement of high $p_T$ heavy hadron yields observed in experiments~\cite{Cao:2018ews, Scardina:2017ipo, ALICE:2015vxz, Wiedemann:2000tf}. This phenomenon is primarily attributed to the energy loss or gain experienced by high-energy heavy quarks as they traverse the medium created post-collision. During their passage through the QGP medium, heavy quarks interact with the medium constituents, resulting in either energy loss or gain, depending on the nature of the interaction. These interactions are broadly categorized into two classes: (i) hard collisions with plasma constituents, which include elastic (\(2 \rightarrow 2\)) and inelastic collisions or the emission of gluonic radiation (\(2 \rightarrow 3\)), and (ii) soft collisions due to interactions with plasma collective excitations or soft modes. In hard collisions, the momentum transfer is high, on the order of the medium temperature (\(T\)). In contrast, in soft collisions, the momentum transfer is low, on the order of the Debye mass, \(m_D \sim g_s T\), where \(g_s\) is the strong running coupling. 

The study of heavy quark energy loss is well-covered in existing research, mainly focusing on elastic and inelastic collisions or radiation~\cite{Svetitsky:1987gq, Moore:2004tg, GolamMustafa:1997id, Burke:2013yra, Adil:2006ei, Dusling:2009jn, Cho:2009ze, Elias:2014hua, Dokshitzer:2001zm, Zhang:2003wk, Kurian:2020orp, Kumar:2021goi, Nezhad:2023mvb, Mustafa:2004dr, Qin:2007rn, Peng:2024zvf}. This review  focuses on a relatively underexplored aspect: the energy loss of heavy quarks resulting from their interaction with soft collective modes in the QGP, commonly called the polarization energy loss. Additionally, we highlight the existing gaps in the literature surrounding this phenomenon and suggest areas for further investigation. The study of collective modes has been addressed in several works considering various plasma aspects~\cite{Mrowczynski:2005ki, Romatschke:2003ms, Romatschke:2004jh, Schenke:2006yp, Schenke:2006xu, Carrington:2014bla, Jamal:2017dqs, Jamal:2019pxr, YousufJamal:2018ucf, Kumar:2017bja}, revealing a diverse spectrum similar to electromagnetic plasmas~\cite{Blaizot:2001nr}. These modes can be accurately described in classical fields, offering valuable insights into the temporal evolution of a non-equilibrium system moving towards equilibrium. The characteristics of these modes are highly dependent on various medium properties, such as anisotropy, chemical potential, and non-ideal or quasi-particle effects. Consequently, 
these medium properties significantly influence the interaction between heavy quarks and the collective modes. Therefore, to study polarization energy loss, two major areas require understanding: first, the interaction between heavy quarks and the QGP medium, particularly with the collective modes generated within the medium, and second, the modelling of the medium itself, which involves accounting for non-ideal effects or the quasi-particle nature, the non-equilibrium and anisotropic characteristics of the QGP medium, and the presence of finite chemical potential. It is important to note that factors such as magnetic fields, viscosity, and chirality are not included in this review as they are beyond the scope of the current discussion.

Before going deeper into our primary aim, exploring the various facets of studying the QGP medium is crucial. Developing reliable and versatile theoretical tools for the quantitative analysis of QCD plasmas—both in and out of equilibrium—is essential as our understanding of QCD plasma dynamics continues to evolve despite significant advancements in recent years. The commonly employed theoretical frameworks include transport or effective kinetic theory, diagrammatic techniques such as quantum field theory (notably, hard thermal loop (HTL) perturbation theory), lattice simulations, and various hydrodynamic models. Each methodology offers unique benefits and faces specific limitations, making them suited for addressing different aspects of QCD plasma behaviour. For instance, HTL perturbation theory provides an extensive framework for understanding plasma behaviour. However, it is challenged by non-perturbative aspects of long-wavelength excitations, particularly in weakly coupled plasmas nearing equilibrium. Conversely, while lattice simulations have been successful for non-perturbative studies of static properties, they encounter difficulties with dynamic scenarios. Traditional Monte Carlo simulations of QCD at finite quark chemical potential face challenges~\cite{Litim:2001db}. In contrast, kinetic theory has emerged as an efficient method for calculating macroscopic properties such as transport coefficients. 

Despite its limitations, kinetic theory is extensively applied due to its flexibility in dealing with both equilibrium and non-equilibrium systems~\cite{Blaizot:2001nr}. This review follows the kinetic theory approach, which articulates the time-evolving distribution of particles within phase space defined by their momenta and positions. These distribution functions change due to inter-particle collisions and interactions with external or self-consistently generated mean fields. We will review the formalism used in various articles, considering different plasma aspects. Realizing the importance of the non-ideal behaviour of the medium particles, we have a dedicated section with a brief overview and comparison of widely used quasi-particle models, highlighting their strengths and weaknesses.

The review is organized as follows: In Section~\ref{sec:formalism}, we discuss the methodology for analyzing collective modes and polarization energy loss within the kinetic theory approach, considering factors such as medium particle collisions, momentum anisotropy, and finite chemical potential. We also discuss the instabilities that significantly influence plasma dynamics. Section~\ref{sec:QP} is dedicated to quasi-particle models, while Section~\ref{sec:EG} focuses on the possibility of energy gain by heavy quarks through soft contributions. Section~\ref{sec:EX} discusses the experimental aspects of the current study, and Section~\ref{sec:Sum} addresses the summary and possible future directions.

\section{COLLECTIVE MODES AND THE POLARIZATION ENERGY LOSS}
\label{sec:formalism}
This review is centred on the polarization energy loss experienced by heavy quarks as they interact with soft collective modes in the QGP medium. Several research groups have investigated these modes and polarization energy loss separately or together under different conditions. For instance, studies on isotropic or equilibrium QGP mediums include works such as Refs.~\cite{Klimov:1982bv, Jamal:2017dqs, Jamal:2020fxo}, thoroughly examined the relevant phenomena. The effects of anisotropy have been explored in several studies, including Refs.~\cite{Romatschke:2003ms, Romatschke:2004au, Carrington:2014bla, Carrington:2015xca, Zhao:2023mrz, Mrowczynski:2016etf, Carrington:2011uj}, highlighting how anisotropic conditions influence them. Moreover, the impact of medium particle collisions has been addressed in studies like Refs.~\cite{Kumar:2017bja, Schenke:2006xu}, while the effects of finite chemical potential have been explored in Refs.~\cite{Jamal:2020emj, Jamal:2022ztl}. Additionally, several studies have employed a quasi-particle description to analyze these processes, including Refs.~\cite{YousufJamal:2019pen, Jamal:2017dqs, Kumar:2017bja}. This section will first outline the general formalism that is the foundation of these analyses. Following this, we will systematically review the results of these studies in various subsections, paying particular attention to the different aspects of the interactions of heavy quarks with collective modes in the QGP medium, along with the gaps of studies in the literature.

The general mechanism of the collective modes is as follows. We start by considering the QGP in a homogeneous and stationary state, with no net local colour charges or currents. As a heavy quark moves through the QGP, it disturbs the medium, generating local charges and currents. These disturbances create colour fields, which interact with the particles in the medium, inducing further charges and currents. This interaction strengthens the medium's collective response, leading to the formation of collective modes. The quest for collective excitations in the context of QGP commenced around 1981 ~\cite{Klimov:1982bv}, marking the initiation of a research trajectory that has seen substantial advancements by subsequent authors \cite{Romatschke:2003ms, Romatschke:2004au, Schenke:2006xu, Carrington:2014bla, Jamal:2017dqs, Zhao:2023mrz}. These modes can be either quark or gluonic. However, the main focus remained on gluonic modes due to their central role in strong interactions. Also, the soft gluonic modes in plasma are highly occupied due to the Bose-Einstein distribution in the weak coupling limit $g_s < 1$. While moving in the QGP medium, the heavy quarks interact with these modes. A general condition for energy gain or loss is as follows: if the heavy quark's velocity is greater than the phase velocity of the propagating mode or, equivalently, the phase of the induced colour field, the heavy quark transfers energy to the modes, resulting in energy loss. Conversely, if the heavy quark's velocity is less than the phase velocity, it gains energy. When the heavy quark's velocity equals the phase velocity, the probabilities of energy loss and gain are equal. We will discuss it under different scenarios and provide brief derivations whenever needed.

To analyze polarization energy loss, one describes the dynamics of heavy quarks using the Wong equations, treating it as a classical particle moving in the presence of field within the $SU(N_c)$ gauge group ~\cite{Wong:1970fu}. The Wong equations in Lorentz covariant form are \cite{Jamal:2020fxo, Carrington:2015xca}:

\ba
\frac{dX^{\mu}(\tau)}{d\tau} &=& V^{\mu}(\tau), \label{eq:1_1} \\
\frac{dP^{\mu}(\tau)}{d\tau} &=& g_s Q^{a}(\tau)F^{\mu\nu}_{a}(X(\tau))V_{\nu}(\tau), \label{eq:1_2} \\
\frac{dQ^{a}(\tau)}{d\tau} &=& -g_s f^{abc}V_{\mu}(\tau)A^{\mu}_{b}(X(\tau))Q_{c}(\tau). \label{eq:wong}
\ea
Here, $\tau$ denotes proper time, and $X^\mu (\tau)$, $V^\mu (\tau)$, and $P^\mu (\tau)$ represent the position, velocity, and momentum of the heavy quark, respectively. The parameters include $Q_a$ (colour charge), $g_s$ (strong coupling), $F^{\mu\nu}$ (chromodynamic field tensor), $A^\mu$ (gauge potential), $f^{abc}$ (structure constant of $SU(N_c)$), and $N_c$ (number of colour charges). To solve these equations, the two key assumptions are employed \cite{Carrington:2012fnq, YousufJamal:2019pen}:
\begin{enumerate}
    \item The energy lost per unit time by the particle is small compared to the particle’s total energy, allows to neglect changes in velocity during motion.
    \item The gauge condition $V_{\mu}A^{\mu}_{a}(X) = 0$ implies that the gauge potential vanishes along the heavy quark’s trajectory, rendering the right-hand side of Eq.~\eqref{eq:wong} to zero, so $Q^{a}$ remains constant.
\end{enumerate}
Focusing on the $\mu=0$ component of Eq.~\eqref{eq:1_2} and replacing proper time $\tau$ with time $t=\gamma\tau$, the energy loss rate can be derived as:
\ba
\frac{dE}{dt} &= g_s Q^a {\bf v} \cdot {\bf E}_{a}(t, {\bf x}={\bf v}t), \label{eq:el1st}
\ea
where ${\bf E}_{a}(t,{\bf x}={\bf v}t)$ is the chromoelectric field induced by the heavy quark's moving with velocity, ${\bf v} = {\bf p}/E$ having energy, $E = \sqrt{p^2 + M^2}$. Eq.\eqref{eq:el1st}  can also be expressed in terms of the external current, ${\bf j}^a_{ext} (t, {\bf x})= g_{s} Q^a {\bf v} \delta^{(3)}({\bf x} - {\bf v}t)$ associated with the heavy quark:

\ba
\frac{dE}{dt} &= \int d^3{\bf x}\, {\bf E}^{a}(t, {\bf x}) \cdot {\bf j}^a_{ext} (t, {\bf x}).
\label{eq:el1}
\ea

To determine the form of \({\textbf E}^{a}\), one can employ the linearized Yang-Mills approach and apply the standard procedure of Fourier transforming the linearized differential equation into its algebraic form:

\ba
-i k_{\nu}F^{\nu\mu}_{a}(K) = j^{\mu}_{a, ind}(K) + j^{\mu}_{a, ext}(K),
\label{Maxwell:1}
\ea
where \(K^\mu=({\omega,{\bf k}}) \equiv K\) and the induced current $j^{\mu}_{a, ind}(K)$ is given by:

\ba
j^{\mu}_{a, ind}(K) = \Pi^{\mu\nu}(K) A_{a, \nu}(K),
\label{eq:LIC}
\ea
where $\Pi^{\mu\nu}(K)$ is the gluon selfenergy. Eq.\eqref{Maxwell:1} with Eq.\eqref{eq:LIC} can be rearranged as:
\ba
[K^{2}g^{\mu\nu} - k^{\mu}k^{\nu} + \Pi^{\mu\nu}(K)]A_{a, \nu}(K) = -j^{\mu}_{a, ext}(K).
\label{eq:eom}
\ea
In the temporal gauge $A_{0} = 0$ and $A^j_a = {\text E}^{i}_a(K)/i\omega$ one can obtained the induced chromo-electric field as:
\begin{align}
{\text E}^{i}_a(K)=i\omega \Delta^{ij}(K)j^{j}_{a, ext}(K).
\label{eq:eind}
\end{align}
Hence, the induced chromo-electric field can be expressed in terms of the external current (\(j^{j}_{a, ext}(K)\)) and gluon propagator \(\Delta^{ij}(K)\)~\cite{Thoma:1990fm, Chakraborty:2006md}. The general expression for them in the Fourier space are given as follows~\cite{YousufJamal:2019pen}:
\ba
\Delta^{ij}(K) &=& \left[ (|{\bf k}|^2 - \omega^2)\delta^{ij} - k^i k^j + \Pi^{ij}(K) \right]^{-1}, \label{eq:jind10}
\ea
\ba
j^{j}_{a, ext}(K) &=& \frac{i g_s q^a v^j}{\omega - {\bf k} \cdot {\bf v} + i \epsilon},
\label{eq:jind1}
\ea
where $\epsilon$ is the infinitesimal quantity and  $``i\epsilon"$ prescription makes the Fourier transformed vanish for $t < 0$. The external current described in Eq.~\eqref{eq:jind1} remains unaffected by the medium; however, the propagator, which includes \(\Pi^{ij}(K)\), accounts for medium effects. The dispersion relation for the collective modes can be derived from the poles of this propagator. By substituting Eq.~\eqref{eq:jind10} and Eq.~\eqref{eq:jind1} into Eq.~\eqref{eq:eind}, the induced chromo-electric field can be obtained, which determines the variation in heavy quark energy in Fourier space. To return to coordinate space, an inverse Fourier transform is further needed. {\it  Importantly, \(\Pi^{ij}(K)\) plays a crucial role in both the behaviour of collective modes and polarization energy loss, making it an essential quantity to study under various conditions.} As previously mentioned, \(\Pi^{ij}(K)\) has been calculated in different scenarios by various authors, which will be systematically reviewed in the following subsections. We will begin to review with the simplest case, i.e., isotropic QGP medium, and move towards the more complex phenomena.

\subsubsection{The isotropic QGP medium}
\label{sec:iso}
We begin with a brief discussion of the derivation of gluon selfenergy in the isotropic medium and then review the results from various articles. The heavy quark motion disturbs the stationary state of the QGP medium, which leads to a small change in the distribution functions of the medium particles, denoted by $\delta f^{i}_a(q, X)$, where $X=x^{\mu}=(t,{\bf x})$ is the space-time four-vector, and $q$ is the momentum of the medium particles. The index $``i"$ represents the medium particle species (quark, anti-quark and gluon). The resulting distribution function, $f^{i}_a(q,X)$, can be expressed in a linear approximation as \cite{Kumar:2017bja}:
\begin{align}
    f^{i}_a(q,X)=f^{i}({\bf q})+\delta f^{i}_a(q, X),
\end{align}
where $f^{i}({\bf q})$  is the isotropic distribution function and $\delta f^{i}_a(q, X)$ is small compared to $f^{i}({\bf q})$ but contains crucial information about medium response towards the cause of disturbance or perturbation. Due to a change in the colour charge particle distribution, the induced current $j_{\text{ind}, a}^{\mu}(X)$ arises from each species $``i"$ present in the medium, which can be obtained as \cite{Mrowczynski:2007hb}:
\begin{align}
    \label{eq:jind}
    j_{\text{ind},a}^{\mu}(X) &= g_s\int\frac{d^{3}q}{(2\pi)^3} {u}^{\mu}\Big(2N_c \delta f^{g}_a(q,X) + N_{f}[\delta f^{q}_a(q,X) -\delta f^{\bar{q}}_a(q,X)]\Big),
\end{align}
where $N_{f}$ is the number of quark flavors. $U=u^{\mu}=(1,{\bf u}) = q^\mu/E_q$ is the velocity four-vector with $q^\mu = (E_q, {\bf q})$, the four-momentum and $m$ is the mass the medium particles. To obtain the unknown, $\delta f$, one needs to solve the Boltzmann equation ~\cite{10.1088/978-0-7503-1200-4ch1} that describes the evolution of a single-particle distribution function using a collision term, $\mathcal{C}^{i}_a(q,X)$ that includes the microscopic interactions of the system given as \cite{Kumar:2017bja}:
\begin{align}
    \label{eq:BM}
    {u}^{\mu }\partial^{x}_{\mu}\delta f_a^i(q,X) + g_s\theta _{i} {u}_{\mu }F^{\mu \nu }_a(X)\partial^q _{\nu }f^{i}({\bf q}) = \mathcal{C}^{i}_a(q,X),
\end{align}
where $F_{a}^{\mu\nu}(X)$ is the chromoelectromagnetic strength tensor. The parameter $\theta _{i}$ takes the value $1$ for quarks and gluons and $-1$ for anti-quarks. The challenge in solving Eq.~\eqref{eq:BM} stems from the non-linearity of the collision term, as it includes the product of distribution functions ~\cite{Mitra:2018akk}. Therefore, several approximations are made to linearize the theory, and a few kernels are defined that we shall discuss in section\ref{sec:col}. For the collisionless case, $\mathcal{C}^{i}_a(q,X) = 0$, solving Eq.\eqref{eq:BM} for each $\delta f_a^i$ and inserting it into Eq.\eqref{eq:jind} gives the induced current in the Fourier space as~\cite{Schenke:2006xu}:
\begin{align}
 j^{\mu}_{a, ind}(K) &= g_s^2 \int \frac{d^3 q}{(2\pi)^3} u^{\mu} \partial^{\beta}_{(q)} f({\bf q}) \bigg[ g_{\alpha \beta} - \frac{u_{\alpha} k_{\beta}}{k\cdot u + i \epsilon}\bigg] A^{\alpha}_a(K).
\label{eq:induced current}
\end{align}
The distribution function $f({\bf q})$ is the combination of quark, anti-quark, and gluon distribution functions given as:
\begin{align}
\label{fq}
f({\bf q}) = 2 N_c f_{g}({\bf q})+ N_{f} (f_{q}({\bf q})+f_{\bar{q}}({\bf q})),
\end{align}
where $f_{q,g} ({\bf q})$ denote the quark, antiquark and gluon distribution functions. In the absence of finite chemical potential, $f_{q} ({\bf q})= f_{{\bar q}} ({\bf q})$. For the ideal case when the particles are not interacting with the medium, i.e., no thermal interaction, the isotropic/equilibrium distribution functions reduce to the Bose-Einstein for gluons and Fermi-Dirac for quarks given as\cite{Blaizot:2001nr}:
 \ba
    f_{g} = \frac{\exp[-E_g/T]}{1- \exp[- E_g/T]}, ~~~
    f_{q,{\bar q}} = \frac{\exp[-E_q/T]}{1+ \exp[- E_q/T]},
    \label{eq:iso_dist}
\ea
 where $E_g = |{\bf q}|$ is the energy of gluons and $E_q = \sqrt{|{\bf q}|^2+m^2}$ is the energy of light quarks and anti-quarks present in the medium. Since $m\ll T$, one may neglect the mass of the light quarks, and hence, the energies of medium particles are effectively the same. In the linear approximation, using the equation of motion for the gauge field, $A_{\nu}(X)$, given in Eq.\eqref{eq:LIC}, one can extract the $\Pi^{\mu\nu}(K)$ for the isotropic QGP medium as\cite{Schenke:2006xu}:
\begin{align}
   \Pi^{\mu\nu}(K) &= g_s^{2}\int \frac{d^{3}q}{(2\pi)^{3}} u^{\mu}\frac{\partial f({\bf q})}{\partial q^{\beta}}\bigg[g^{\nu\beta} - \frac{u^{\nu}k^{\beta}}{k \cdot u + i\epsilon}\bigg],
   \label{iso_pi}
\end{align}
where $\Pi^{\mu\nu}(K) = \Pi^{\nu\mu}(K)$ and adheres to Ward's identity, $k_{\mu}\Pi^{\mu\nu}(K) = 0$. In other words, $\Pi^{\mu\nu}(K)$ is symmetric and transverse in nature, which guarantees that the current given by Eq.\eqref{eq:induced current} is gauge-independent. Hence, one can choose to work with spatial coordinates. Next, the gluon selfenergy for the isotropic medium given in Eq.\eqref{iso_pi} has one independent vector, {\it i.e., {\bf k}}, which can be decomposed into two independent components, usually chosen to be transverse, $P^{ij}_{T}(K)$ and longitudinal, $P^{ij}_{L}(K)$, as \cite{Romatschke:2003ms, Carrington:2014bla}:
\begin{align}
\Pi^{ij}(K) = P^{ij}_{T}(K) \Pi_{T}(K) + P^{ij}_{L}(K) \Pi_{L}(K),
\label{eq:pinu}
\end{align}
where \(P^{ij}_{T}(K) = \delta^{ij} - k^{i}k^{j}/{k^{2}}\) and \(P^{ij}_{L}(K) = k^{i}k^{j}/{k^{2}}\). On solving Eq.\eqref{iso_pi} and Eq.\eqref{eq:pinu} simultaneously with proper contraction one can obtain \cite{Romatschke:2003ms, Blaizot:2001nr}:
\begin{align}
\Pi_T(K) & = m_D^2\frac{\omega}{4 k^3} \bigg[2 \omega k + \Big(k^2-\omega^2\Big)\ln \Big(\frac{\omega+k }{\omega -k }\Big)\bigg],
\label{eq:pt}
\end{align}
and
\begin{align}
\Pi_L(K) =-m_D^2\frac{\omega ^2}{k^2}\left(1-\frac{\omega }{2 k}\ln \left(\frac{\omega +k }{\omega -k }\right)\right).
\label{eq:pl}
\end{align}
These structure functions exhibit logarithmic terms that will have complex results as:
\begin{equation}
\ln\frac{\omega+k}{\omega-k} = \ln\frac{|\omega+k|}{|\omega-k|} - i \pi \Theta(k^2-\omega^2),
\label{eq:reimen}
\end{equation}
where $\Theta$ is the heavy-side theta function, which ensures that the imaginary part only contributes when $k^2>\omega^2$ or the phase velocity, $v_p = \omega/k <1$. Next, the Debye mass, \(m_D\), contain the information of the isotropic distribution functions as \cite{Carrington:2014bla}:
\begin{align}
m_D^2 &= -4 \pi \alpha_{s}(T) \bigg(2 N_c \int \frac{d^3 q}{(2 \pi)^3} \partial_q f_g ({\bf q})+2 N_f \int \frac{d^3 q}{(2 \pi)^3} \partial_q f_q ({\bf q})\bigg),
\label{dm}
\end{align}
where \(\alpha_{s}(T)\) is the QCD running coupling constant at the finite temperature given as\cite{Laine:2005ai},
\ba
\alpha_{s}(T) = \frac{g_{s}(T)^2}{4\pi}.
\label{eq:alp}
\ea
From Eq.\eqref{eq:iso_dist} and Eq.\eqref{dm}, with Eq.\eqref{eq:alp}, the final expression for the Debye mass becomes\cite{Carrington:2014bla}:
\ba
    \label{massd}
    m_D^2 &=& \frac{g_s^2 T^2}{3}\left(N_c + \frac{N_f}{2}\right).
\ea
Now, using the gluon selfenergy for isotropic QGP medium, one can have the gluon propagator and obtain the polarization energy loss of heavy quarks using Eq.\eqref{eq:el1}. After averaging over colour indices and performing \(\omega\)-integration using the Residue theorem (accounting for the constrain imposed by the external current at \(\omega = {\bf k} \cdot {\bf v}\)), Eq.\eqref{eq:el1} can be simplified by \cite{Jamal:2020emj}:
\begin{align}
    -\frac{dE}{dt} = \frac{\alpha_s C_F}{2 \pi^2}
    \int d^3{\bf k} \frac{\omega}{|{\bf k}|^2} \left\{ \omega^2 \text{Im}\left[ \frac{1}{-\omega^2 + \Pi_L} \right] + \left(|{\bf k}|^2 |{\bf v}|^2 - \omega^2 \right) \text{Im}\left[ \frac{1}{-\omega^2 + k^2 + \Pi_T} \right] \right\}_{\omega = {\bf k} \cdot {\bf v}},
    \label{eq:de_general}
\end{align}
where \(C_F\) denotes the Casimir invariant of \(SU(N_c)\). From Eq.\eqref{eq:de_general}, one can conclude that if the structure functions of gluon selfenergy are purely real, the energy loss is zero. Only the complex (especially imaginary) solution contributes to the energy loss. This is the main result of this subsection. Alternatively, in some of the articles, the energy loss is expressed in terms of the longitudinal (\(\epsilon_L(K)\)) and transverse (\(\epsilon_T(K)\)) dielectric functions \cite{Jiang:2014oxa, Jamal:2020fxo, Mrowczynski:2007hb}:
\begin{align}
    -\frac{dE}{dt} = \frac{\alpha_s C_F}{2 \pi^2 |{\bf v}|^2} \int^{k_{\text{max}}}_{0} d^3{\bf k} \frac{\omega}{k^2} \left\{ \left(k^2 |{\bf v}|^2 - \omega^2\right) \text{Im} \left[ \frac{1}{\omega^2 \epsilon_T(K) - k^2} \right] + \text{Im} \left[ \frac{1}{\epsilon_L(K)} \right] \right\}_{\omega = {\bf k} \cdot {\bf v}},
    \label{eq:de_dielectric}
\end{align}
which can be easily seen by using the following relations \cite{Blaizot:2001nr}:
\begin{align}
    \epsilon_L(K) = 1 - \frac{\Pi_L(K)}{\omega^2}, \quad
    \epsilon_T(K) = 1 - \frac{\Pi_T(K)}{\omega^2}.
    \label{eq:dielectric_def}
\end{align}
The physical interpretation of Eq.\eqref{eq:de_dielectric} is as follows: 
\paragraph{Interaction with the collective modes:}
The poles of the integrand in Eq.~\eqref{eq:de_dielectric} correspond to the collective modes of the plasma. Let us call them A-mode and G-mode. Their dispersion relations are, respectively, determined by \cite{Carrington:2015xca, Blaizot:2001nr}:
\begin{align}
    \epsilon_L(K) = 0, \quad \epsilon_T(K) = \frac{k^2}{\omega^2}.
\label{eq:disp}
\end{align}
Here, A- mode describes the behaviour of longitudinal modes (plasmons), while G- mode captures the dynamics of transverse modes. Therefore, physically, Eq.\eqref{eq:de_dielectric} shows that the heavy quark does not interact with plasma constituents but rather with the plasma collective modes. Hence, the systematic study of these collective modes helps us understand the energy loss of heavy quarks as they traverse the QGP.

\paragraph{ Ultraviolet Divergence and Cutoff:}
The energy loss expression provided in Eq.~\eqref{eq:de_dielectric} is incomplete due to a logarithmic divergence as \(k_{\text{max}} \rightarrow \infty\). This occurs because Eq.~\eqref{eq:de_dielectric} is derived using a classical approximation, which becomes invalid for large \(k\), leading to ultraviolet divergence. To resolve this issue, an upper cutoff, \(k_{\text{max}}\), must be introduced. 
\begin{itemize}
    \item In Refs. \cite{Jiang:2014oxa, Jiang:2016duz, Jamal:2020fxo}, the cutoff is chosen as \(k_{\text{max}} \sim m_D \sim g_s T\), representing interactions with soft modes, which are relevant to the current discussion.
    \item To obtain the total collisional energy loss, Eq.~\eqref{eq:de_dielectric} must be supplemented with the hard contribution, which describes the elastic collisions of a test parton with plasma constituents involving momentum transfers significantly greater than the Debye mass. Unlike the soft contribution, the hard contribution does not exhibit ultraviolet divergence, as the maximum momentum transfer is limited by collision kinematics. In Refs. \cite{Chakraborty:2006db, Debnath:2023zet}, the cutoff for momentum transfer, \(k_{\text{max}}\), is taken as the maximum momentum exchange in a \(2 \rightarrow 2\) collision between the heavy quark and medium particles (or hard modes), resulting in a momentum-dependent expression:
\end{itemize}
\begin{equation}
    k_{\text{max}} = \min \left[E, \frac{2q(E + p)}{\sqrt{2q(E + p) + M^2}}\right],
    \label{eq:kmax}
\end{equation}
where \(E = \sqrt{p^2 + M^2}\) is the energy of the heavy quark, and \(q \sim T\) denotes the typical momentum of the medium particles in the QGP. 

In Ref. \cite{Carrington:2015xca}, it is demonstrated that increasing \(k_{\text{max}}\) results as increase in energy loss per unit time.
In the first scenario, as shown in Ref.
\cite{YousufJamal:2019pen}, the results show saturation in the high-momentum limit (\(p \gg M\)) because, at high momentum, Eq.~\eqref{eq:de_dielectric} becomes independent of \(p\), leading to a plateau in energy loss. In contrast, in the second scenario, as found in Ref.\cite{Chakraborty:2006db}, the energy loss expression retains its dependence on \(p\) and continues to increase with rising momentum. As noted in Ref. \cite{Mrowczynski:2007hb}, the cutoff beyond \(k_{\text{max}} \sim m_D\) is associated with hard collisions involving plasma constituents, which are not relevant to the current analysis. 
\begin{figure}[ht]
		\begin{center}
			\includegraphics[height=6cm,width=8cm]{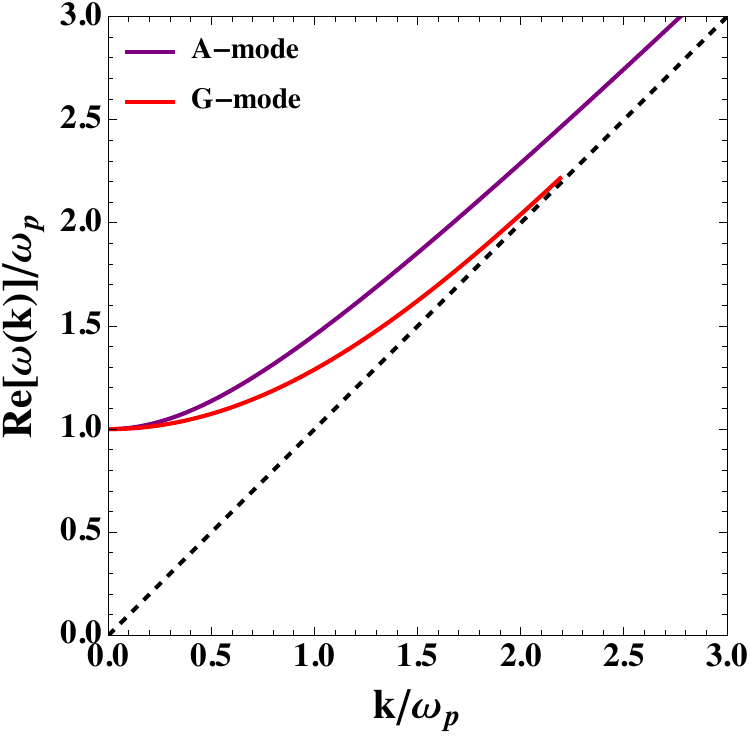}
			\caption{The real part of collective modes dispersion relations within the isotropic QGP medium at temperature, T = 0.25 GeV ~\cite{Jamal:2017dqs}.}
			\label{modes_iso}
		\end{center}
	\end{figure}

The dispersion relations can produce complex values for \(\omega(k)\), where stable modes exist if \(\text{Im}(\omega(k)) = 0\). If \(\text{Im}(\omega(k)) > 0\), the mode becomes unstable and grows exponentially, while \(\text{Im}(\omega(k)) < 0\) indicates damping. Determining the exact number of solutions for each dispersion equation can be challenging due to the presence of logarithmic terms. When approximations are applied, there is a risk of either missing a solution or generating a spurious one as an artifact of the approximation. Similarly, when using numerical methods, the search is often confined to a specified range, potentially overlooking solutions outside of that range. A Nyquist analysis, however, can be used to determine the number of solutions to a given equation without explicitly solving it. It is essential to note that our focus in this review is solely on discussing the types of modes under different scenarios and how they will affect polarization energy loss. Therefore, we avoid much discussion of Nyquist analysis. For further details, readers are encouraged to refer to Refs.~\cite{Carrington:2014bla, Carrington:2021bnk} and the references therein. In Ref. \cite{Carrington:2014bla} through Nyquist analysis, it is claimed that there are six real modes possible in the isotropic medium, where the longitudinal mode (A-mode) and transverse mode (G-modes) are illustrated in Fig.\ref{modes_iso}.  Here, $\omega_p = m_D/\sqrt{3}$ denotes the minimum frequency of the plasma wave when $k = 0$ and is referred to as the plasma frequency. In Refs. \cite{Romatschke:2003ms, Romatschke:2004au, Carrington:2015xca, Jamal:2017dqs, Kumar:2017bja}, it is concluded that no damping and unstable modes exist in the isotropic medium. As a result, the only contribution to the energy loss comes from the pole at \(\omega = {\bf k} \cdot {\bf v}\), which originates from the heavy quark's current.
\begin{figure}[ht]
\begin{center}
    \includegraphics[height=6cm,width=8cm]{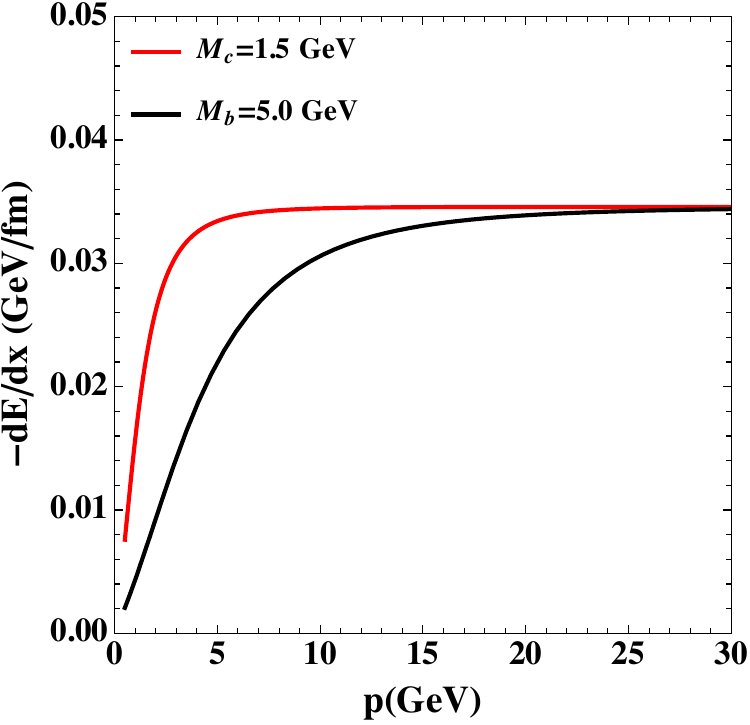} 
    \caption{Comparison of the energy loss per unit length, $-dE/dx$ of a charm quark, M = 1.5 GeV and bottom quark, M = 5.0 GeV as a function of momenta, $p$ for T = 0.3 GeV in the isotropic hot QGP medium~\cite{YousufJamal:2019pen}.}
    \label{fig_pol_kmax_md}
    \end{center}
    \end{figure}
    
The energy loss expression in Eq.~\eqref{eq:de_dielectric} is real and finite only when either \(\epsilon_L(K)\) or \(\epsilon_T(K)\) has a non-zero imaginary component. According to Eq.~\eqref{eq:reimen}, this imaginary part arises only when the phase velocity of the modes is less than unity. Due to the condition \(\omega = {\bf k} \cdot {\bf v}\), where \(\omega < k\) for \(0 < v < 1\), the imaginary component consistently contributes to the polarization energy loss. The results for the energy loss per unit length, \(\frac{dE}{dx} = \frac{1}{v} \frac{dE}{dt}\), from Ref. \cite{YousufJamal:2019pen} are shown in Fig.~\ref{fig_pol_kmax_md} as a function of the heavy quark's momentum. Here, authors have employed Eq.~\eqref{eq:de_dielectric}, where \(v = p/\sqrt{p^2+M^2}\). At higher quark momentum, the energy loss tends to saturate due to the soft contribution. For heavier quarks, such as the bottom quark, saturation occurs more gradually than the charm quark. Likewise, Ref. \cite{Jiang:2014oxa} presents comparable results for an isotropic, non-viscous QGP medium. Similar results are presented in Ref. \cite{Jamal:2020fxo}, where the authors extended \(k_{\text{max}}\) up to \(2 m_D\). The small differences in magnitude between these studies are due to variations in parameters like heavy quark masses and medium temperature. In Ref. \cite{Jiang:2014oxa}, the authors also show results for a massless test parton with velocity \(|{\bf v}| \rightarrow 1\), which represents the upper limit for energy loss in the polarization case. After saturation, the energy loss for the charm and bottom quarks converges with the massless case, as they behave like massless particles at extremely high momentum.  These were the results for the simplest scenario, the isotropic QGP medium. In the following subsection, we will review the impact of finite collisions among the medium particles on the collective modes and polarization energy loss.

\subsubsection{Presence of medium particle collision}
\label{sec:col}
In the previous subsection, we reviewed studies based on the collisionless case, where the Boltzmann equation, Eq.\eqref{eq:BM}, is solved with the collision term \(\mathcal{C}^{i}_a(q,X) = 0\), meaning that the mean free path of the medium particles is larger than the system's lifetime, i.e., the QGP medium. However, when the medium is slightly away from equilibrium and the particles have a finite collision frequency, these collisions must be considered in the analysis. To account for the microscopic interactions within the QGP medium, the collision term \(\mathcal{C}^{i}_a(q,X)\) is included in Eq.\eqref{eq:BM}, transforming the Boltzmann equation into an integro-differential equation, making the problem more complex. Several well-established kernels have been proposed to model these interactions effectively for simplification. Among these, the Relaxation Time Approximation (RTA) kernel where all interactions are governed by a relaxation time~\cite{Anderson:1974nyl}, the Bhatnagar-Gross-Krook (BGK) kernel~\cite{Bhatnagar:1954zz}, and others \cite{Bleicher:2022kcu, BOERCKER197943, Sorensen:2023zkk} are effective in describing systems near equilibrium. However, extra conditions are needed in the RTA to conserve the energy, momentum, and particle number. On the other hand, the BGK kernel is more general and becomes the same as RTA in some instances. The BGK kernel naturally conserves particle number, so only one condition is needed to ensure energy and momentum conservation, as explained in Ref.\cite{Singha:2024mkk}. Since most studies on collective modes and polarization energy loss use the BGK kernel, we will focus on this method and review related results from various research articles.

The BGK collisional kernel, introduced by Bhatnagar, Gross, and Krook in 1954, simplifies the complex collision terms in the Boltzmann equation while conserving particle numbers for each collision. It effectively simulates the impact of binary collisions characterized by significant momentum transfer. Understanding how these collisions influence the response of the QGP medium is crucial. The mathematical structure of the BGK kernel is expressed as~\cite{YousufJamal:2019pen}:
\begin{equation}
\mathcal{C}^{i}_a(q,X) = -\nu\left[f^{i}_a(q,X) - \frac{N^{i}_a(X)}{N^{i}_{\text{eq}}}f^{i}_{\text{eq}}(|\mathbf{q}|)\right],
\label{eq:BGK}
\end{equation}
where \(N^{i}_a(X)\) represents the particle number density, and \(N^{i}_{\text{eq}}\) is its equilibrium value. The BGK kernel signifies the system's equilibration due to collisions in a time proportional to \(\nu^{-1}\), where $\nu$ is the collision frequency, often considered as a constant parameter. Although this assumption may not hold universally, it is widely employed in various studies \cite{Kumar:2017bja, Jiang:2016dkf, Han:2017nfz}. The RTA kernel can be recovered from Eq.\eqref{eq:BGK} in the limit \(\frac{N^{i}_a(X)}{N^{i}_{\text{eq}}} \rightarrow 1\). As mentioned above, the BGK kernel conserves particle numbers instantaneously, a feature that the RTA kernel does not hold. This conservation is expressed as:

\begin{equation}
\int \frac{d^{3}q}{(2\pi)^3}\mathcal{C}^{i}_a(q,X) = 0,
\end{equation}
implying that collisions occur without altering the total particle number, although the momentum distribution of particles may change. The BGK kernel also conserves the colour current, as demonstrated in Ref.~\cite{Manuel:2004gk}. In Ref. \cite{Singha:2023eia} it is shown that to ensure energy and momentum conservation in the Boltzmann equation with the BGK collision kernel, one requires the condition \( \epsilon n_0 = \epsilon_0 n \),  which is identified as a necessary matching condition. Here, $\epsilon$ and $n$ represent the energy density and net-number density, respectively, whereas $\epsilon_0$ and $n_0$ correspond to their equilibrium value.  Next, one can solve Eq.\eqref{eq:BM} with the BGK kernel, to yield the induced current as~\cite{Schenke:2006xu}:

\begin{align}
j^{\mu \,a}_{\text{ind}}(K) &= g_s^2 \int\frac{d^{3}q}{(2\pi)^3}u^{\mu}\partial_{\nu}^{(q)}
f(\mathbf{q})\mathcal{M}^{\nu\alpha}(K,U)D^{-1}(K,\mathbf{u},\nu)A_{\alpha}^{a} 
 +  g_s i \nu \Bigg\{2 N_c \mathcal{S}^{g}(K,\nu) \nn
& + N_f \left(\mathcal{S}^{q}(K,\nu) - \mathcal{S}^{\bar{q}}(K,\nu)\right)\Bigg\}  + g_s i \nu \int \frac{d\Omega}{4\pi}u^{\mu}D^{-1}(K,\mathbf{u},\nu) \int\frac{d^{3}q^{\prime}}{(2\pi)^3} \Bigg[g_s \partial_{\nu}^{(q^{\prime})} f(\mathbf{q^{\prime}})  \nn 
& \times \mathcal{M}^{\nu\alpha}(K,u^{\prime})D^{-1}(K,\mathbf{u^{\prime}},\nu)\mathcal{W}^{-1}(K,\nu)
A_{\alpha}^{a} +i\nu\left(f_{\text{eq}}(|\mathbf{q^{\prime}}|) - f(\mathbf{q^{\prime}})\right)D^{-1}(K,\mathbf{u^{\prime}},\nu)\Bigg] \mathcal{W}^{-1}(K,\nu),
\label{fullcurrent}
\end{align}
where $f(\mathbf{q})$ is given in Eq.\eqref{fq} and
\begin{equation}
f_{\text{eq}}(|\mathbf{q^{\prime}}|) = 2N_c f_{\text{eq}}^{g}(|\mathbf{q^{\prime}}|) + N_f\left[f_{\text{eq}}^q(|\mathbf{q^{\prime}}|) + f_{\text{eq}}^{\bar{q}}(|\mathbf{q^{\prime}}|)\right],
\end{equation}

\begin{equation}
D(K,\mathbf{u},\nu) = \omega + i\nu - \mathbf{k} \cdot \mathbf{u},
\end{equation}

\begin{equation}
\mathcal{M}^{\nu\alpha}(K,U) = g^{\nu\alpha}(\omega - \mathbf{k} \cdot \mathbf{u}) - K^{\nu}u^{\alpha},
\end{equation}

\begin{equation}
\mathcal{W}(K,\nu) = 1 - i \nu \int \frac{d\Omega}{4\pi}D^{-1}(K,\mathbf{u},\nu),
\end{equation}

\begin{equation}
\mathcal{S}^{i}(K,\nu) = \theta_i\int\frac{d^{3}q}{(2\pi)^3} u^{\mu}[f^{i}(\mathbf{q}) - f^{i}_{\text{eq}}(|\mathbf{q}|)]D^{-1}(K,\mathbf{u},\nu).
\end{equation}
From Eq.\eqref{eq:LIC}, one can extract the gluon selfenergy in the presence of particle collisions in the medium as \cite{Jamal:2020emj}:
\begin{align}
\Pi^{\mu\nu}(K) &= g_s^2 \int\frac{d^{3}q}{(2\pi)^3} u^{\mu} \partial_{\beta}^{(q)}f(\mathbf{q}) \mathcal{M}^{\beta\nu}(K,U) D^{-1}(K,\mathbf{u},\nu) + \delta_{ab} g_s^2 (i \nu) \int \frac{d\Omega}{4\pi} u^{\mu} \nn
& \times D^{-1}(K,\mathbf{u},\nu) \int\frac{d^{3}q^{\prime}}{(2\pi)^3} \partial_{\beta}^{(q^{\prime})} f(\mathbf{q}^{\prime}) \mathcal{M}^{\beta\nu}(K,u^{\prime}) D^{-1}(K,\mathbf{u}^{\prime},\nu) \mathcal{W}^{-1}(K,\nu).
\label{selfenergy}
\end{align}
After further simplification and using the temporal gauge, the selfenergy can be written as \cite{Jamal:2020emj}:
\begin{align}
\Pi^{ij}(K) &= m_D^2 \int \frac{d\Omega}{4 \pi} u^{i} u^{l} \left(\frac{u^{j} k^{l} + \left(\omega - {\bf k} \cdot {\bf u}\right) \delta^{lj}}{\omega + i \nu - {\bf k} \cdot {\bf u}}\right).
\label{eq:pi}
\end{align}
\begin{figure}
		\begin{center}
			\includegraphics[height=7cm,width=7cm]{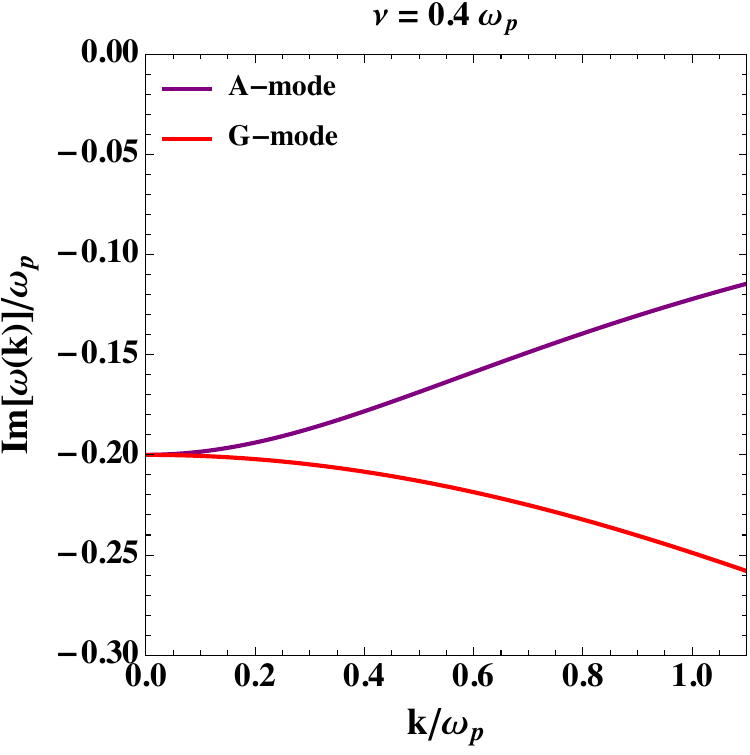}
			\caption{The imaginary solutions of dispersion equations at finite collision frequency in the isotropic QGP medium at T = 0.25 GeV and $\nu$ = 0.4 $\omega_p$ \cite{Kumar:2017bja}.}
			\label{IM_modes_iso_nu}
		\end{center}
	\end{figure}
To obtain the exact form of $\Pi^{ij}(K)$ in the presence of finite collision in the medium, one needs to solve simultaneously, Eq.\eqref{eq:pinu} and \eqref{eq:pi}. Doing so, the following forms of the transverse, $\tilde{\Pi}_T(K)$ and longitudinal, $\tilde{\Pi}_L(K)$ structure functions can be obtained\cite{Jamal:2020emj, Blaizot:2001nr}:
\begin{align}
\tilde{\Pi}_T(K) &= m_D^2 \frac{\omega}{4 k^3} \bigg[2 k (\omega + i \nu) + \Big(k^2 + (\nu - i \omega)^2\Big) \ln \Big(\frac{\omega + i \nu + k}{\omega + i \nu - k}\Big) \bigg],
\label{eq:ptnu}
\end{align}
and
\begin{align}
\tilde{\Pi}_L(K) &= -m_D^2 \frac{\omega^2}{k^2} \left(\frac{1 - \frac{\omega + i \nu}{2 k} \ln \left(\frac{\omega + i \nu + k}{\omega + i \nu - k}\right)}{1 - \frac{i \nu}{2 k} \ln \left(\frac{\omega + i \nu + k}{\omega + i \nu - k}\right)}\right).
\label{eq:plnu}
\end{align}

\begin{figure}
		\begin{center}
	\includegraphics[height=6cm,width=6cm]{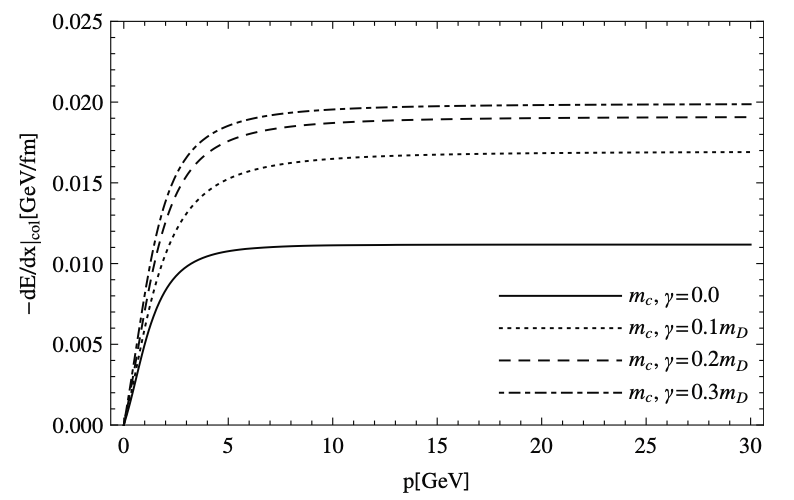} 
        \includegraphics[height=6cm,width=6cm]{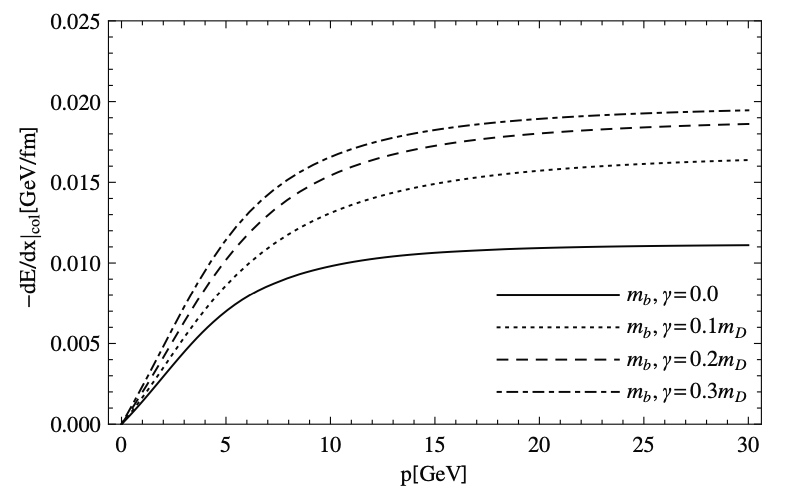} 
\caption{Comparison of polarization energy loss per unit length at different collision frequencies, $\gamma \equiv\nu$ plotted as a function of momentum, $p$. The charm quark mass, M = 1.5 GeV (left panel), and bottom quark mass, M = 5 GeV (right panel) at T = 0.3 GeV ~\cite{Han:2017nfz}.}
			\label{fig_pol_1}
		\end{center}
	\end{figure}
For detailed derivations, see Refs.~\cite{Jamal:2020emj, Schenke:2006xu, Kumar:2017bja, Jiang:2016dkf, Han:2017nfz}. When the constituent particles undergo collisions, no additional vector is needed for the analysis. Consequently, we retain both the dielectric functions, \(\tilde{\epsilon}_L(K)\) and \(\tilde{\epsilon}_T(K)\), both modified by the effects of medium particle collisions. This again results in two dispersion relations, as given in Eq.~\eqref{eq:disp}. However, due to the presence of an additional \(i\) term in the structure functions, the solution of the dispersion relations becomes complex with a negative imaginary part corresponding to the damping of these types of modes \cite{Schenke:2006xu}. Hence, the magnitude of the previously mentioned real modes is suppressed. Moreover, as illustrated in Fig.~\ref{IM_modes_iso_nu}, two new imaginary modes emerge for a collision frequency \(\nu = 0.4 \omega_p\) as shown in Refs.\cite{Kumar:2017bja, Jamal:2022ztl}. As per the previous discussion, these imaginary modes fall under the category of damped modes, yet they do not qualify as over-damped. As the collision frequency increases, the system moves farther from equilibrium. Significantly, these modes contribute further to the energy loss of the heavy quark. It is because the Logarithmic term in the present case expands as \cite{Jiang:2016dkf}:
\begin{equation}
\ln\frac{\omega+k+i\nu}{\omega-k+i\nu} = \ln\frac{\sqrt{(\omega^2 - k^2 + \nu^2)^2 + 4k^2\nu^2}}{(\omega - k)^2 + \nu^2} - i \arccos\left(\frac{\omega^2 - k^2 + \nu^2}{\sqrt{(\omega^2 - k^2 + \nu^2)^2 + 4k^2\nu^2}}\right),
\label{eq:log_nu}
\end{equation}
where the imaginary part contains the effect of the collision frequency. From Eq.\eqref{eq:log_nu} with Eq.\eqref{eq:ptnu} and \eqref{eq:plnu}, it is evident that the imaginary part of the dielectric functions and hence, the gluon propagator depends on the collision frequency. Ref. \cite{Han:2017nfz} highlights an important point: when considering collisional energy loss, \(-\frac{dE}{dx}\) is proportional to the cross-section. In a robust estimation, the transport cross-section \(\sigma_{tr}\) is related to the collision rate as \(\sigma_{tr} \propto \nu\). Therefore, an increase in the collision rate naturally results in an increase in \(\sigma_{tr}\), leading to more significant collisional energy loss. This crucial point warrants further exploration in the literature, as similar findings are observed in Fig.~\ref{fig_pol_1}, adapted from Ref. \cite{Han:2017nfz}. The collisions between medium particles enhance energy loss, with higher collision frequencies causing greater losses. This indicates that when the system is away from the equilibrium, the heavy quark traversing through it experiences increased energy loss. In Ref. \cite{YousufJamal:2019pen}, the authors compare the polarization energy loss of charm and bottom quarks using the RTA and BGK kernels, keeping the collision frequency constant. They report that, for the same momentum and collision frequency, the energy loss is greater in the RTA scenario than in the BGK. Additionally, the bottom quark experiences less energy loss than the charm quark under similar conditions. It would be interesting to see further studies that adopt more advanced methods to account for medium particle collisions. Next, we will review the case of anisotropy without considering collisions, followed by a discussion on the anisotropic scenario with finite collisions.

\subsubsection{Anisotropic QGP medium}
\label{sec:aniso}

In off-central HICs, the initial geometry of the system approximates an almond shape, leading to spatial anisotropy. As a result, distinct pressure gradients in different directions manifest as momentum anisotropy in the final state. While the QGP eventually reaches local equilibrium after a brief period \cite{Bozek:2010aj, Bozek:2022cjj}, the momentum distribution of the pre-equilibrium plasma remains anisotropic. Despite its short-lived nature, this anisotropic phase significantly impacts the collective modes and, hence, the energy loss of heavy quarks. The weakly coupled anisotropic QGP is inherently unstable due to the spontaneous growth of chromodynamic fields. The medium particle distribution functions capture the anisotropy. Romatschke and Strickland, in 2003 \cite{Romatschke:2003ms} proposed an elegant Ansatz for parameterizing an anisotropic momentum distribution by deforming the isotropic counterparts along a preferred direction as:
\begin{align}
f({\mathbf{q}}) \rightarrow f_{\xi}({\mathbf{q}}) = C_{\xi} ~ f\left(\sqrt{{\bf q}^2 + \xi ({\bf q} \cdot {\bf \hat{n}})^2}\right),
\label{eq:disan}
\end{align}
where ${\bf \hat{n}}$ is a unit vector representing the anisotropic direction, typically chosen along the beam direction. The parameter $\xi \in (-1,\infty)$ controls the distribution's shape, quantifying the anisotropic strength in the medium. It describes the degree of squeezing ($\xi > 0$, oblate form) or stretching ($-1 < \xi < 0$, prolate form) along the ${\bf \hat{n}}$ direction, with $\xi = 0$ yielding the isotropic case. In several investigations, $\xi$ is treated as a free parameter. A more realistic portrayal of $\xi$ is defined in Ref. \cite{Zhang:2020efz} as:
\begin{align}
\xi = \frac{\langle {\bf q_T^2} \rangle}{2 \langle q_L^2 \rangle} - 1,
\end{align}
where the longitudinal momentum component is defined as \( q_L = \mathbf{q} \cdot \mathbf{\hat{n}} \), and the transverse momentum component is given by \( \mathbf{q}_T = \mathbf{q} - \mathbf{\hat{n}} (\mathbf{q} \cdot \mathbf{\hat{n}}) \). For small anisotropy, in Ref.\cite{Asakawa:2006jn}, the parameter \(\xi\) is linked to the shear viscosity to entropy density ratio (\(\eta/s\)) of the medium via the one-dimensional Bjorken expansion in the Navier-Stokes limit as:
\begin{align}
\xi = \frac{10}{T \tau} \frac{\eta}{s},
\end{align}
where \(1/\tau\) represents the expansion rate of the fluid element. Furthermore, the normalization constant \(C_\xi\) given in Eq. \eqref{eq:disan} was initially set to unity \cite{Romatschke:2003ms}. However, Romatschke and Strickland \cite{Romatschke:2004au} normalized it based on the anisotropic number density relative to the isotropic one, yielding \(C_\xi = \sqrt{1 + \xi}\). An alternative expression incorporating the normalized Debye mass, \(m_D\), is provided by Carrington et al. \cite{Carrington:2014bla}:
\begin{align}
C_{\xi} = 
\begin{cases}
\frac{\sqrt{|\xi|}}{\tanh^{-1} \sqrt{|\xi|}} & \text{if } -1 \leq \xi < 0, \\
\frac{\sqrt{\xi}}{\tan^{-1} \sqrt{\xi}} & \text{if } \xi \geq 0.
\end{cases}
\end{align}

Now, to obtain the gluon selfenergy in the anisotropic QGP medium, one can follow the methodology mentioned in the isotropic case, employing the anisotropic distribution function from Eq.\eqref{eq:disan}. Doing so, the following expression can be obtained \cite{Kumar:2017bja}:
\begin{align}
\Pi^{ij}(K) &= m_D^2 C_{\xi} \int \frac{d\Omega}{4 \pi} u^i \frac{u^l + \xi (\mathbf{u} \cdot \mathbf{\hat{n}}) n^l}{(1 + \xi (\mathbf{u} \cdot \mathbf{\hat{n}})^2)^2} \Big[\delta^{jl} + \frac{u^j k^l}{k^\mu u_\mu + i \epsilon}\Big].
\label{aniso_pi}
\end{align}

In this anisotropic scenario, an additional vector (\({\bf \hat{n}}\)) is introduced. Therefore, two additional projections are needed to accommodate this vector: one to represent the direction of \({\bf \hat{n}}\) and another to relate the two vectors \({\bf \hat{n}}\) and \({\bf k}\) as done in Ref.\cite{Kobes:1990dc}. One possible form for these projections could be \cite{Romatschke:2003ms, Kumar:2017bja}:
\ba
P^{ij}_{n} = \frac{{\tilde{n}}^i {\tilde{n}}^j}{{\tilde{n}}^2}, \quad P^{ij}_{kn} = k^{i} {\tilde{n}}^{j} + k^{j} {\tilde{n}}^{i},
\ea
where \(\tilde{n}^i = (\delta^{ij} - \frac{k^i k^j}{k^2}) \hat{n}^j\) is a vector orthogonal to \(\mathbf{k}\), i.e., \(\mathbf{\tilde{n}} \cdot \mathbf{k} = 0\). The gluon selfenergy can be decomposed as follows:
\ba
\Pi^{ij} = \alpha P^{ij}_{T} + \beta P^{ij}_{L} + \gamma P^{ij}_{n} + \delta P^{ij}_{kn},
\label{seexpan}
\ea
where \(\alpha\), \(\beta\), \(\gamma\), and \(\delta\) are structure functions that can be determined by the following contractions of \(\Pi^{ij}\):
\ba
\begin{aligned}
    \alpha &= (P^{ij}_{T} - P^{ij}_{n}) \Pi^{ij}, \quad \beta = P^{ij}_{L} \Pi^{ij}, \\
    \gamma &= (2 P^{ij}_{n} - P^{ij}_{T}) \Pi^{ij}, \quad \delta = \frac{1}{2 k^{2} {\tilde{n}}^2} P^{ij}_{kn} \Pi^{ij}.
    \label{structurefunctions}
\end{aligned}
\ea

In the small anisotropy limit, utilizing equations \eqref{aniso_pi}, \eqref{seexpan}, and \eqref{structurefunctions}, these structure functions can be analytically derived \cite{Carrington:2014bla}:
\ba
\alpha &=& C_{\xi} m_D^2 \bigg[\frac{z^2}{2} + \frac{\xi }{48 k} \bigg\{ \bigg(-2 k \left(15 z^4 - 19 z^2 + 4\right) \cos (2 \theta_n) \nn &-& 2 k \left(9 z^4 - 9 z^2 + 4\right)\bigg) + \bigg(\bigg(3 k \left(z^2 - 1\right) \left(5 z^2 - 3\right) z \cos (2 \theta_n) \nn
&-& 3 k \left(z^2 - 1\right) z + 9 k \left(z^2 - 1\right) z^3\bigg) - \frac{1}{4} z \left(z^2 - 1\right)\bigg) \log \left(\frac{z + 1}{z - 1}\right)\bigg\}\bigg],
\label{eq:axi}
\ea
\ba
\beta &= C_{\xi} \bigg[ m_D^2 \left(-z^2\right) \bigg(\xi \cos^2(\theta_n) - \frac{2 \xi}{3} - \frac{1}{2} \xi z^2 (3 \cos (2 \theta_n) + 1) \nn
& + \frac{1}{4} z \log \left(\frac{z + 1}{z - 1}\right) \left(\xi \left(3 z^2 - 2\right) \cos (2 \theta_n) + \xi z^2 - 2 \right) + 1 \bigg)\bigg],
\label{eq:bxi}
\ea
\ba
\gamma &=& C_{\xi} \frac{m_D^2}{12} \xi \left(z^2 - 1\right) \sin^2(\theta_n) \bigg(6 z^2 - 4 - 3 \left(z^2 - 1\right) z \log \left(\frac{z + 1}{z - 1}\right)\bigg),
\label{eq:cxi}
\ea
\ba
\delta &=& C_{\xi} \frac{m_D^2 z \cos{\theta_n}}{48 k} \xi \Bigg(\left(88 z - 96 z^3\right) + \Big(12 \left(4 z^4 - 5 z^2 + 1\right)\log \frac{z + 1}{z - 1}\Bigg),
\label{eq:dxi}
\ea
where \(z = \frac{\omega}{k}\). These results are valid for any angle \(\theta_n\), where \(\theta_n\) is the angle between \(\mathbf{\hat{n}}\) and \(\mathbf{k}\) and small anisotropy ($\xi<1$) \cite{Kumar:2017bja}.
In Ref.\cite{Schenke:2006xu, Romatschke:2004au}, the authors numerically calculated these structure functions for large anisotropy (\(\xi > 1\)), and the analytical solutions are obtainable only for \(\mathbf{\hat{n}} \parallel \mathbf{k}\) and \(\mathbf{\hat{n}} \perp \mathbf{k}\). Putting the gluon selfenergy given in Eq.\eqref{seexpan} in the propagator, $[\Delta(K)]^{ij}$ given in Eq.\eqref{eq:jind10} one finds:
\begin{equation}
[\Delta^{-1}(K)]^{ij}=(k^{2}-{\omega}^{2}+\alpha){P}^{ij}_{T}+(\beta-{\omega}^{2}){P}^{ij}_{L}+\gamma{P}^{ij}_{n}+\delta{P}^{ij}_{kn}.
\label{invpropexpan}
\end{equation}

In order to find the poles of the propagator, $[\Delta(K)]^{ij}$, one needs to know the exact form of $[\Delta(K)]^{ij}$. Next, we will briefly discuss a general way of obtaining the poles of the propagator in the anisotropic case. Owing to the fact that both a tensor and its inverse lie in the same space spanned by some basis vectors (projection operators), $[\Delta(K)]^{ij}$ can also be expanded as its inverse \cite{Mrowczynski:2016etf, Carrington:2014bla}:
\begin{equation}
[\Delta(K)]^{ij}=a{P}^{ij}_{L}+b{P}^{ij}_{T}+c{P}^{ij}_{n}+d{P}^{ij}_{kn}.
\label{propagator}
\end{equation}

Now, using the relation $[\Delta^{-1}(K)]^{ij} [\Delta(K)]^{jl}=\delta^{il}$, one can obtain the expression for the coefficients $a$, $b$, $c$, and $d$ as done in Ref. \cite{Kumar:2017bja, Romatschke:2003ms, Schenke:2006xu}. Doing so, one can yield the following form of the propagator given in Eq.\eqref{propagator} as:
\ba
[\Delta(K)]^{ij}&=&\Delta_A({P}^{ij}_{T}-{P}^{ij}_{n})+\Delta_G\big[(k^2-\omega^2+\alpha+\gamma){P}^{ij}_{L}+ (\beta-\omega^2){P}^{ij}_{n}-\delta {P}^{ij}_{kn}\big].
\label{propagator1}
\ea
Hence, the poles can only be obtained from $\Delta_{A}(K)$ and $\Delta_{G}(K)$. Their inverse can be obtained as follows \cite{Jamal:2022ztl}:
\begin{equation}
\Delta^{-1}_{A}(K)=k^2-\omega^2+\alpha=0,
\label{mode_a}
\end{equation}
and
\begin{equation}
\Delta^{-1}_{G}(K)=(k^2-\omega^2+\alpha + \gamma)(\beta-\omega^2)-k^2 \tilde n^2 \delta^2=0.
\label{mode_g}
\end{equation}

In the small anisotropy, one can neglect, $\delta^2$ as it will provide a contribution of $\xi^2$ and hence, Eq.\eqref{mode_g} can give two possible solutions as:
\begin{equation}
\Delta_{G1}^{-1}(K)=k^2 - \omega^2 + \alpha + \gamma=0,
\label{mode_g1}
\end{equation}
and
\begin{equation}
\Delta_{G2}^{-1}(K)=\beta-\omega^2=0.
\label{mode_g2}
\end{equation}

\begin{figure}[ht]
		\begin{center}
			\includegraphics[height=6cm,width=8cm]{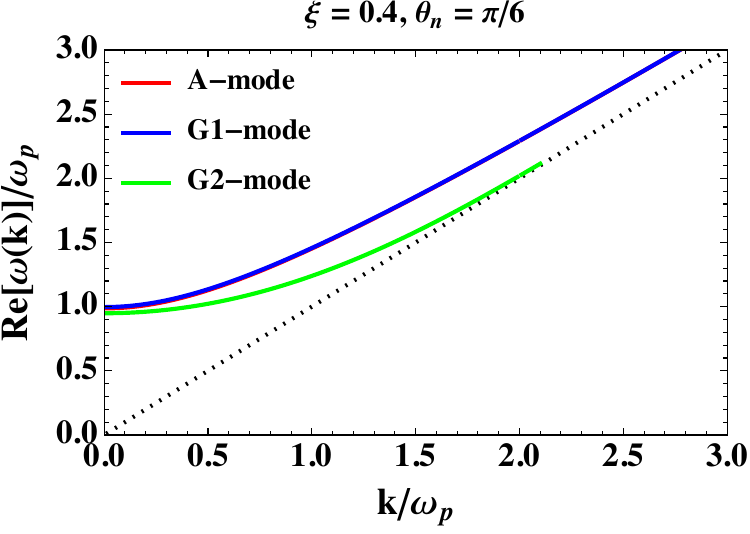}
			\caption{The real part of collective modes dispersion relations within the anisotropic QGP medium at temperature, T = 0.25 GeV~\cite{Kumar:2017bja}.}
			\label{fig:modes_aniso}
		\end{center}
	\end{figure}

\begin{figure}[ht]
		\begin{center}
			\includegraphics[height=6cm,width=8cm]{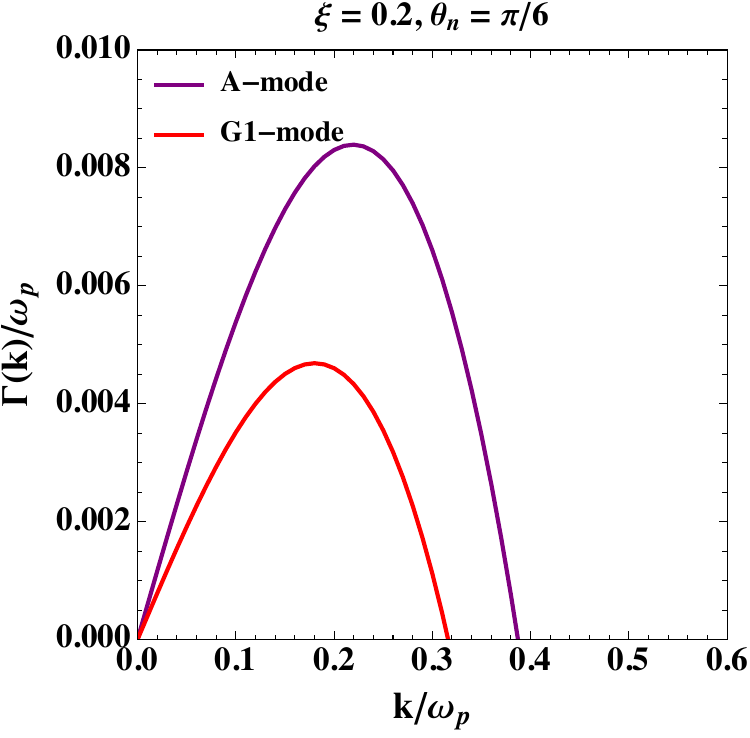}
			\caption{The dispersion relation of unstable modes at small anisotropy, $\xi$ = 0.2, $\theta_n$ = $\pi/6$, temperature, T = 0.25 GeV~\cite{Kumar:2017bja}.}
			\label{modes_aniso}
		\end{center}
	\end{figure}

\begin{figure}[ht]
		\begin{center}
			\includegraphics[height=6cm,width=8cm]{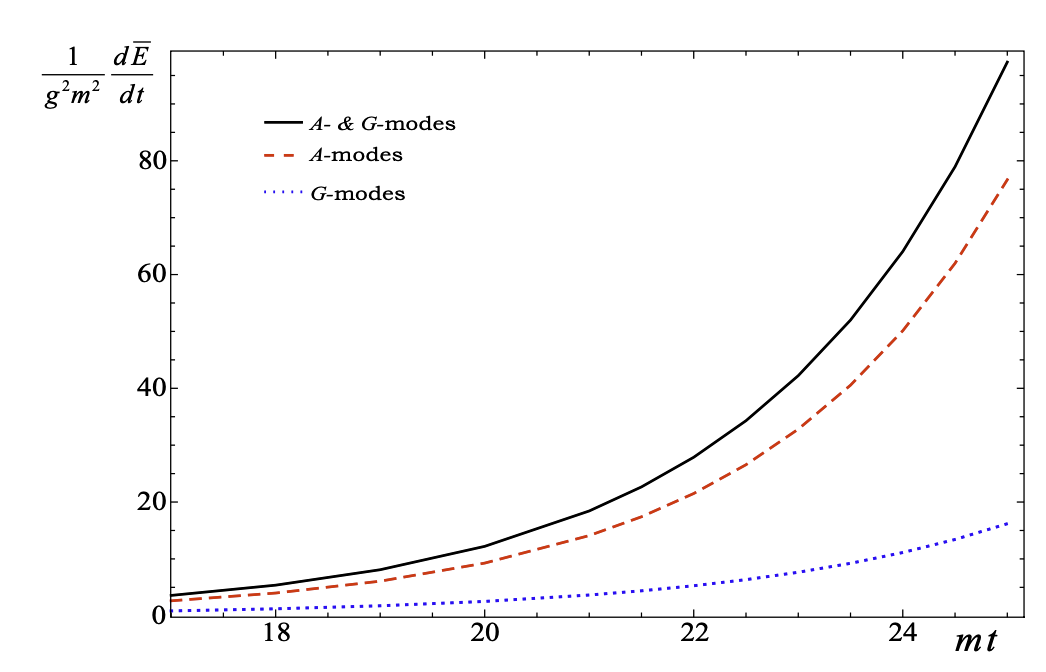}
			\caption{  The energy loss per unit time in oblate plasma as a function of time when the momentum of parton is perpendicular to the anisotropy vector. The red (dashed) line corresponds to the effect of A modes, the blue (dotted) line represents the G modes, and the black (solid) line represents the sum. Here, $m\equiv m_D$ (screening mass) given in Eq.\eqref{dm} and g is the strong coupling~\cite{Carrington:2015xca}.}
			\label{carr_EL}
		\end{center}
	\end{figure}

Again, the Nyquist analysis can be employed to determine the possible number of solutions \cite{Carrington:2014bla}. In the anisotropic case, one gets three dispersion equations (\ref{mode_a}), (\ref{mode_g1}), and (\ref{mode_g2}). In Ref. \cite{Carrington:2014bla}, the authors applied Nyquist analysis to calculate the maximum number of modes, finding eight modes in the prolate case and ten modes in the oblate case. Different authors have assigned various names to these modes; we refer to them as the A-, G1-, and G2- mode dispersion equations, respectively. In the isotropic limit, $\xi\rightarrow 0$,  A- and G1- modes merge as A- mode, and G2- mode becomes G mode. In Ref.\cite{Kumar:2017bja}, the real solutions of their dispersion relations as depicted in Fig.\ref{fig:modes_aniso}. One can notice that A- and G1- modes almost overlap due to minimal contributions from the structure-function $\gamma$.
Besides the real modes, introducing momentum anisotropy in the plasma system leads to additional unstable modes, which significantly influence the plasma dynamics and cause kinetic instabilities beyond contributing to polarization energy loss. In Ref.\cite{Weibel:1959zz, Romatschke:2005ag, Randrup:2003cw, Mrowczynski:1993qm, Mrowczynski:1996vh, Mrowczynski:2004kv, Carrington:2021bnk} these are claimed similar to the Weibel or filamentation instabilities observed in electromagnetic plasma which significantly contribute to swift thermalization processes. These instabilities appear as collective modes with a positive imaginary component of the frequency, $\text{Im}(\omega(k))$. To analyze these modes, we substitute the frequency $\omega$ with a purely imaginary value, $\omega = i \Gamma$. In this case, the structure-function $\beta$ remains positive, so the condition $\beta + \Gamma^2 = 0$ from Eq.~\eqref{mode_g2} is not satisfied ~\cite{Kumar:2017bja}. Hence, only two dispersion equations remain for unstable modes. We illustrate the effects of these instabilities by showing results without considering medium particle collisions in Fig.~\ref{modes_aniso}, at small anisotropy ($\xi = 0.2$) and an angle of $\theta_n = \pi/6$ between the momentum of the medium particles and the direction of anisotropy. The next subsection will discuss the incorporation of medium particle collisions and anisotropy. In Refs.~\cite{Carrington:2014bla, Romatschke:2003ms}, it is shown that these modes are highly dependent on $\theta_n$. However, the G1- mode is completely suppressed at $\theta_n = \pi/2$, whereas at $\theta_n = 0$, the G1- mode overlaps with the A- mode. These unstable modes do not persist at large values of $k$ as shown in Fig.\ref{modes_aniso}. 
An interesting study by Carrington et al.~\cite{Carrington:2015xca} examined a more general case of high-energy parton traversing an anisotropic unstable QGP with rapidly growing collective modes. In the detailed study, the authors show that in equilibrium plasmas, the energy loss follows a well-established formula discussed in subsection\ref{sec:iso}  and does not require initial conditions. However, in the unstable QGP, the energy loss becomes time- and direction-dependent and is influenced by initial chromodynamic fields. If these fields are uncorrelated with the parton's colour state, the parton typically loses energy at a rate comparable to an equilibrium plasma. When the parton induces the fields, the energy transfer can either accelerate or decelerate it, with correlated initial conditions leading to exponential energy gain or loss. The possibility of energy gain is discussed in the section\ref{sec:EG}. In Ref.\cite{Carrington:2015xca}, their numerical results for prolate and oblate plasmas show that energy loss is maximized in specific configurations with a significant directional dependence. In Fig.\ref{carr_EL}, one can observe that the energy transfer is primarily affected by the unstable A mode when the momentum of the parton is perpendicular to the anisotropy vector.

An important point that has been mentioned in the article is the study's implications or jet quenching in relativistic heavy-ion collisions, where fluctuations in energy loss are expected, especially when unstable plasma modes are active. However, to reach there, one needs to fill the gap of studies related to radiative loss in this context, as radiative energy loss behaves differently in unstable plasmas and is expected to grow exponentially. Therefore, there is a promising area for further study that combines these effects and helps in fully understanding jet suppression phenomena. Next, we will discuss the presence of momentum anisotropy along with the finite collisions.

\subsubsection{Finite collisions and momentum anisotropy}
\label{sec:aniso_col}
To combine the study of collision effects with the presence of momentum anisotropy, one needs to consider the medium particle distribution functions given in Eq.\eqref{eq:disan} and a proper collision term, $\mathcal{C}^{i}_a(q,X)$ together in the Boltzmann equation \eqref{eq:BM}. Following the same procedure discussed for each individual case, the expressions for the structure functions for the anisotropic QGP medium using BGK collisional kernel are obtained in Refs. \cite{Schenke:2006xu, Kumar:2017bja} as:
\ba
\tilde{\alpha} &=& C_{\xi} \frac{m_D^2}{48 k} \bigg[12 \bigg(\big(\tilde{z}^2 - 1\big) \log \bigg(\frac{\tilde{z} + 1}{\tilde{z} - 1}\bigg) - 2 \tilde{z}\bigg) (i \nu - k \tilde{z}) + \xi \bigg(\cos \big(2 \theta_n\big) \bigg(3 \big(\tilde{z}^2 - 1\big)
\log \bigg(\frac{\tilde{z} + 1}{\tilde{z} - 1}\bigg)\nn & \times & \big(k \tilde{z} \big(5 \tilde{z}^2 - 3\big)
+ i \nu \big(1 - 5 \tilde{z}^2\big)\big) + k \big(-30 \tilde{z}^4 + 38 \tilde{z}^2 - 8\big) + 2 i \nu \tilde{z} \big(15 \tilde{z}^2 - 13\big)\bigg) + 3 \big(\tilde{z}^2 - 1\big) \nn &\times & \log \bigg(\frac{\tilde{z} + 1}{\tilde{z} - 1}\bigg) \big(k \tilde{z} \big(3 \tilde{z}^2 - 1\big) - i \big(\nu + 3 \nu \tilde{z}^2\big)\big) 
- 2 k \big(9 \tilde{z}^4 - 9 \tilde{z}^2 + 4\big) + 6 i \nu \tilde{z} \big(3 \tilde{z}^2 - 1\big)\bigg],
\ea
\ba
\tilde{\beta} &=& -C_{\xi} \frac{m_D^2 \omega}{k^2} \bigg[\frac{2 k \left(1 - \frac{1}{2} \tilde{z} \log \left(\frac{\tilde{z} + 1}{\tilde{z} - 1}\right)\right) (\nu + i k \tilde{z})}{\nu \log \left(\frac{\tilde{z} + 1}{\tilde{z} - 1}\right) + 2 i k} + \frac{2 k \xi (\nu + i k \tilde{z})}{\nu \log \left(\frac{\tilde{z} + 1}{\tilde{z} - 1}\right) + 2 i k} \bigg(\cos^2(\theta_n) + \frac{1}{12} \big(3 \tilde{z} \big(\big(3 \tilde{z}^2 - 2\big) 
\nn &\times& \log \bigg(\frac{\tilde{z} + 1}{\tilde{z} - 1}\bigg) - 6 \tilde{z}\big) \cos \big(2 \theta_n\big) + 3 \tilde{z}^3 \log \bigg(\frac{\tilde{z} + 1}{\tilde{z} - 1}\bigg) + \frac{\tilde{z} \big(2 - 3 \tilde{z}^2\big)}{2} \bigg) \bigg],
\ea
\ba
\tilde{\gamma} &=& -C_{\xi}\frac{ m_D^2}{12 k} \xi  \left(-6 \tilde{z}^2+3 \left(\tilde{z}^2-1\right) \tilde{z} \log \left(\frac{\tilde{z}+1}{\tilde{z}-1}\right)+4\right) \left(k \left(\tilde{z}^2-1\right)-i \nu  \tilde{z}\right) \sin ^2\left(\theta _n\right),
\ea
and 
\ba
\tilde{\delta} &=& C_{\xi}\frac{m_D^2}{24 k^2} \xi \bigg( \frac{(k \tilde{z}-i \nu ) \cos \big(\theta_n\big)}{2 k-i \nu  \log \big(\frac{\tilde{z}+1}{\tilde{z}-1}\big)} \bigg) \Big(k \big(88 \tilde{z}-96 \tilde{z}^3\big) + \log \big(\frac{\tilde{z}+1}{\tilde{z}-1}\big) \Big(12 k \big(4 \tilde{z}^4-5 \tilde{z}^2+1\big) \nn
&-&3 i \nu  \big(4 \tilde{z}^4-5 \tilde{z}^2+1\big)\log \big(\frac{\tilde{z}+1}{\tilde{z}-1}\big)-10 i \nu  \tilde{z}\Big)+8 i \nu  \big(6 \tilde{z}^2-1\big)
\Big), 
\ea
where $\tilde{z} = \frac{\omega + i \nu}{k}$. In the limit $\xi \rightarrow 0$ and $\nu \neq 0$, the parameters $\tilde{\gamma}$ and $\tilde{\delta}$ tend to zero, while $\tilde{\alpha}$ approaches $\Pi_T$ and $\tilde{\beta}$ approaches $\Pi_L$, as given in Eqs.~\eqref{eq:ptnu} and \eqref{eq:plnu}, respectively. Furthermore, if $\nu \rightarrow 0$ and $\xi \neq 0$, the structure functions are determined by Eqs.~\eqref{eq:axi} - \eqref{eq:dxi}. Additionally, in the limit \{$\nu, \xi\} \rightarrow 0$, one can recover the structure functions of the isotropic collisionless medium, as provided in Eqs.~\eqref{eq:pt} and \eqref{eq:pl}. As shown in Ref.\cite{Kumar:2017bja}, in the small anisotropy limit, the dispersion equations \eqref{mode_a}, \eqref{mode_g}, and \eqref{mode_g1} retain their analytically solvable nature, even when having finite collisions.
\begin{figure}[ht]
		\begin{center}
			\includegraphics[height=6cm,width=8cm]{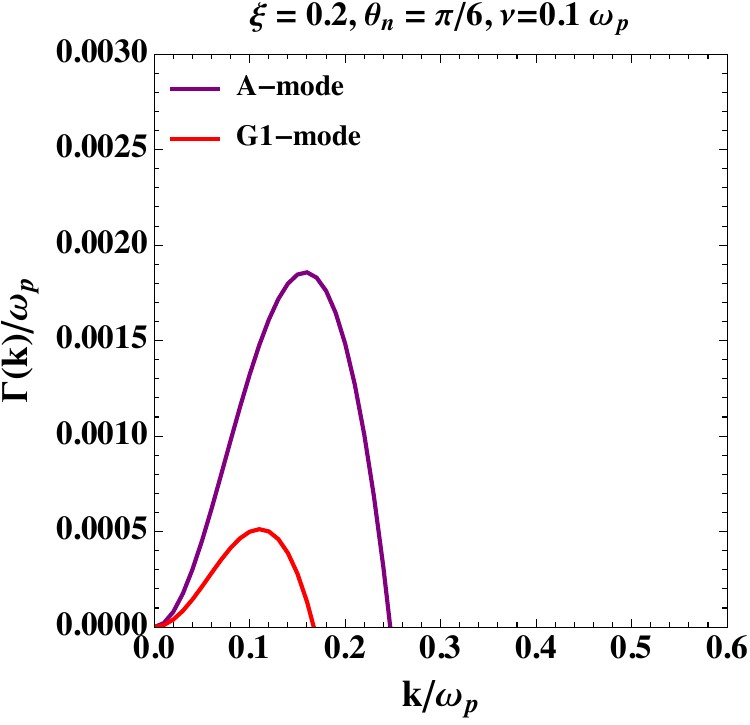}
			\caption{The dispersion relation of unstable modes at small anisotropy, $\xi$ = 0.2, $\theta_n$ = $\pi/6$, temperature, T = 0.25 GeV and collision frequency, $\nu$ = 0.1 $\omega_p$~\cite{Kumar:2017bja}.}
			\label{modes_nu_aniso}
		\end{center}
	\end{figure}
 
\begin{figure}[ht]
		\begin{center}
			\includegraphics[height=6cm,width=7cm]{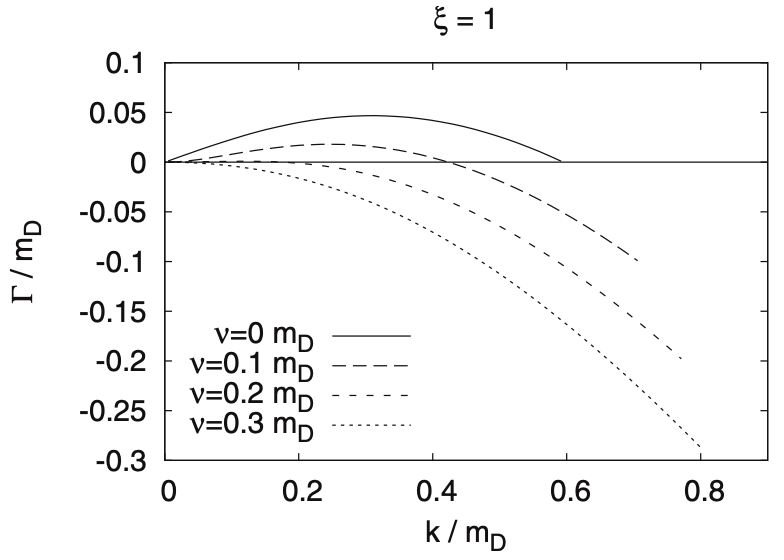}
   \hspace{3mm}
			\includegraphics[height=6cm,width=7cm]{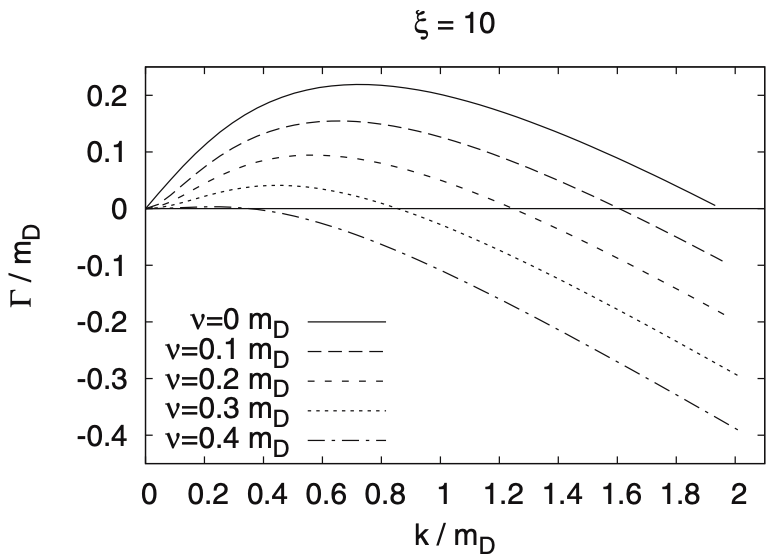}
			\caption{The dispersion relation of unstable modes at $\xi$ = 1 (left panel) and $\xi$ = 10 (right panel) and at different collision frequencies~\cite{Schenke:2006xu}.}
			\label{modes_nu_SLaniso}
		\end{center}
	\end{figure}
 
Since the anisotropy creates extra unstable modes, it is important to understand how these modes are affected by the medium particle collisions and how collectively they modify the energy loss.
The detailed study of collective modes incorporating both momentum anisotropy and medium particle collisions is done in Refs.\cite{Schenke:2006xu, Kumar:2017bja}. Moreover, how these instabilities are suppressed under different scenarios is discussed in great detail. However, as per our current understanding, a combined effect of both on the polarization energy loss is not available in the literature. According to results in Refs.\cite{Schenke:2006xu, Kumar:2017bja} the influence of medium particle collisions significantly suppress these modes, as visually depicted in Fig.~\ref{modes_nu_aniso}, compared to the results shown in Fig.~\ref{modes_aniso}. The magnitudes of the imaginary unstable modes are greatly reduced in the presence of collisions, and they persist only for smaller values of $k$. Notably, even at $\nu = 0.1\omega_p$, these modes experience considerable suppression. Consequently, the complex interplay between medium particle collisions and momentum anisotropy emerges as a crucial factor influencing the evolution of these unstable modes within the QGP medium.

It is to note that in Ref.~\cite{Schenke:2006xu} for computation of structure functions with large anisotropy, $\xi>1$ and finite collisions, the authors focused specifically on the parallel scenario, where \(\theta_n = 0\). On the other hand, Ref.~\cite{Kumar:2017bja} limits its study to small anisotropy, though it provides a quasi-particle description, which will be discussed in section \ref{sec:QP}. In situations of large anisotropy with \(\theta_n = 0\), the structure-function \(\gamma\) vanishes, leaving only the unstable A-mode, as shown in Fig.\ref{modes_nu_SLaniso}.

\begin{figure}[ht]
		\begin{center}
			\includegraphics[height=6cm,width=8cm]{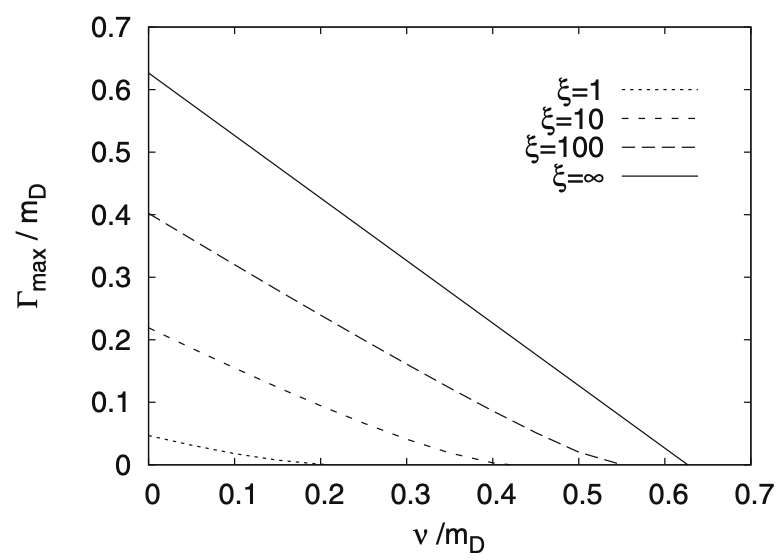}
			\caption{The variation of the maximum strength of unstable modes with respect to the collision frequency at strong anisotropy, $\xi \ge$ 1~\cite{Schenke:2006xu}.}
			\label{schenke_max_xi}
		\end{center}
	\end{figure}
 
In this case, the observed unstable mode is presented in Fig.\ref{modes_nu_SLaniso} for $\xi = 1$ (left panel) and $\xi = 10$ (right panel) where $\Gamma (i \omega)$ and k are normalized by $m_D$. The presence of collisions acts to suppress these instabilities significantly. Notably, they are effectively suppressed at collision frequencies $\nu=0.3 m_D$ and $\nu = 0.4 m_D$, respectively. Furthermore, Ref.~\cite{Schenke:2006xu} clarifies that even under the extreme condition of maximum anisotropy (\(\xi \rightarrow \infty\)), this unstable mode can be fully suppressed at \(\nu \sim 0.6267 m_D \sim 1.0855 \omega_p\), as depicted in Fig.\ref{schenke_max_xi}. Next, as mentioned earlier, a combined study of energy loss with momentum anisotropy and medium particle collision is missing in the literature. It is important to have a study that has a combined effect of both on the energy loss phenomena to get a wider picture. Recent investigations have extended the study of modes to include finite chemical potential while considering small anisotropy and finite collision frequency~\cite{Jamal:2022ztl}. Next, we will discuss the effect of finite chemical potential on the collective modes and polarization energy loss.


\subsubsection{Finite Chemical Potential}
\label{sec:FCP}
Including finite baryon density is essential for studying the QCD phase diagram, especially at low temperatures. High baryon density corresponds to an excess of quarks over antiquarks, characterized by the baryon chemical potential ($\mu_q$). At zero chemical potential, quark and antiquark densities are equal, which is appropriate for high-energy HICs at facilities like RHIC and LHC, and conditions in the early Universe. However, the baryon density becomes an important factor at lower energies, such as those found in the core of neutron stars. This is crucial for experimental facilities operating at moderate temperatures, such as the Super Proton Synchrotron (SPS) at CERN~\cite{Doble:2017syb}, FAIR in Darmstadt, Germany~\cite{Selyuzhenkov:2020djo}, NICA in Dubna, Russia~\cite{Syresin:2022mjz}, and J-PARC in Tōkai, Japan~\cite{Nagamiya:2006en}.

As shown in Fig.~\ref{fig:Phase}, exploring the QCD phase diagram is an important goal in nuclear and particle physics. A critical focus is determining the phase boundary between the QGP and confined phases, particularly at small baryon chemical potentials. Extensive experimental and theoretical efforts have been dedicated to this area~\cite{Fukushima:2010bq, Mohanty:2009vb, Fukushima:2020yzx}. Lattice simulations indicated that the transition between the confined and QGP phases at $\mu_q=0$ is a crossover. There is ongoing debate about whether this crossover could evolve into a first-order transition, implying the existence of a Critical End Point (CEP) in the phase diagram~\cite{Stephanov:2004wx}. Although the precise location and properties of the CEP are still under investigation, different models provide varying predictions for its position in the $T$ vs. $\mu_q$ plane ~\cite{deForcrand:2006pv}. Furthermore, there is uncertainty regarding whether the conjectured phase boundary aligns with the chiral and/or deconfining phase transition line. While initial lattice calculations predicted the existence of the CEP, recent studies have challenged this, noting its absence in some phase diagrams~\cite{deForcrand:2006pv, deForcrand:2008vr}.

Despite these uncertainties, in the context of experiments at moderate temperatures and finite baryon density, theoretical researchers need to study the properties of the hot QCD medium and its various phases at finite chemical potential. A simplified approach to incorporate finite chemical potential into these studies involves modification of the momentum distributions of quarks and antiquarks as follows\cite{Jamal:2022ztl}:
\ba
\label{eq:eq1}
f_{q} &=& \frac{\exp[- (E_q - \mu_q)/T]}{1 + \exp[- (E_q - \mu_q)/T]}, \\
f_{\bar{q}} &=& \frac{\exp[- (E_q + \mu_q)/T]}{1 + \exp[- (E_q + \mu_q)/T]}.
\ea

\begin{figure}[ht]
		\begin{center}
			\includegraphics[height=6cm,width=7cm]{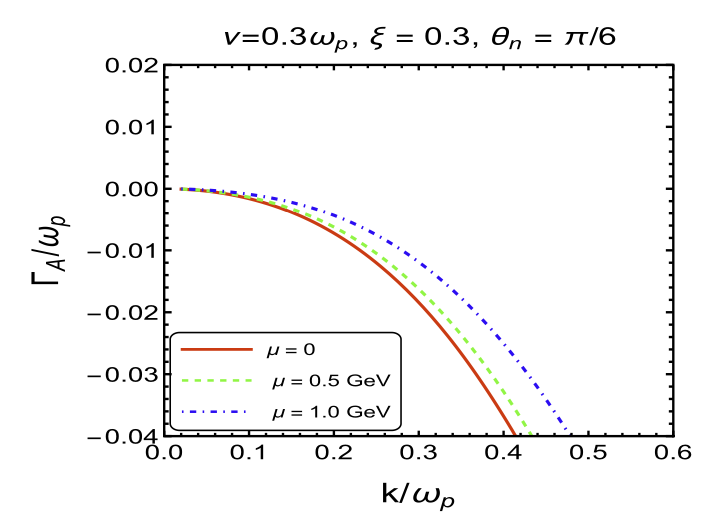}
   \hspace{3mm}
			\includegraphics[height=6cm,width=7cm]{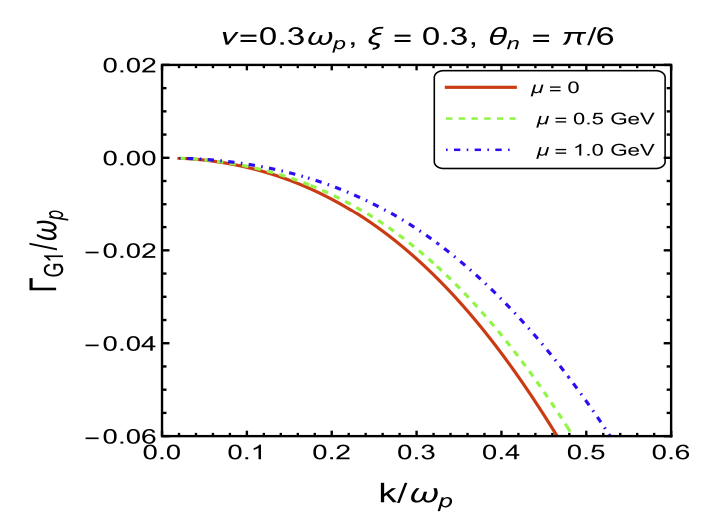}
			\caption{The dispersion relation of unstable A- mode (left panel) and unstable G1- mode (right panel) at $\xi$ = 0.3, $\theta_n = \pi/6$ and $\nu$ =0.3$\omega_p$ at different values of chemical potential, $\mu$ = 0.0, 0.5 and 1.0 GeV~\cite{Jamal:2022ztl}.}
			\label{modes_mu_nu_aniso}
		\end{center}
	\end{figure}

As discussed earlier in Eq.~\eqref{dm}, while calculating the gluon selfenergy, the isotropic distribution function impacts the screening and modifies the Debye screening mass. Hence, using Eq.~\eqref{eq:eq1} with Eq.~\eqref{dm}, the Debye mass $m_D$ modifies as \cite{Jamal:2022ztl}:
\begin{align}
m_D^2(T, \mu_q) &= 4\pi \alpha_s(T, \mu_q) T^2 \bigg[\frac{N_c}{3} + \frac{N_f}{6} + \frac{\mu_q^2}{T^2} \frac{N_f}{2\pi^2} \bigg],
\label{eq:mdh}
\end{align}
where the strong coupling $\alpha_s(T, \mu_q)$ at two-loop order and finite chemical potential is given by~\cite{Kakade:2015laa, Srivastava:2010xa, Bannur:2005wm}:

\begin{figure}
		\begin{center}
	\includegraphics[height=6cm,width=6.5cm]{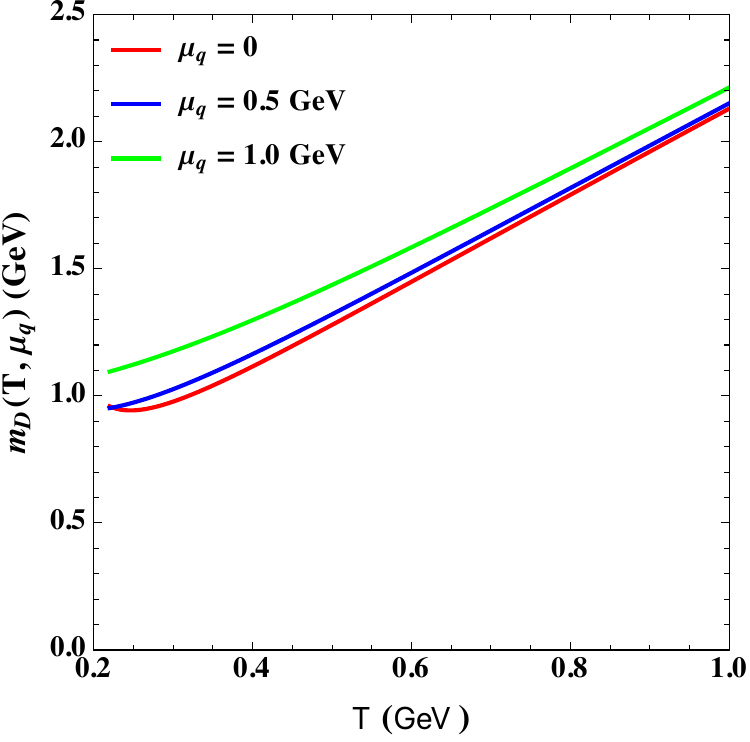} 
 \hspace{3mm}
        \includegraphics[height=6cm,width=6.5cm]{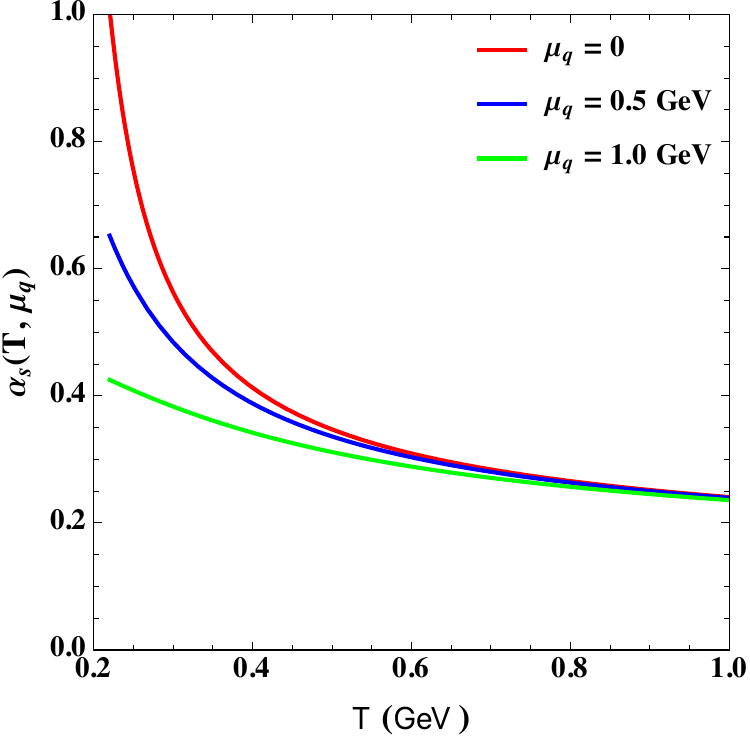} 
\caption{Variation of screening mass (left panel) and strong coupling (right panel) with temperature at different values of chemical potential.}
			\label{fig:mu}
		\end{center}
	\end{figure}
\begin{align}
\alpha_s(T, \mu_q)
&= \frac{6 \pi - \frac{18 \pi (153 - 19 N_f) \ln \left(2 \ln \left(\sqrt{\left(\frac{T}{\Lambda_T}\right)^2 + \frac{\mu_q^2}{\pi^2 \Lambda_T^2}}\right)\right)}{(33 - 2 N_f)^2 \ln \left(\sqrt{\left(\frac{T}{\Lambda_T}\right)^2 + \frac{\mu_q^2}{\pi^2 \Lambda_T^2}}\right)}}{(33 - 2 N_f) \ln \left(\sqrt{\left(\frac{T}{\Lambda_T}\right)^2 + \frac{\mu_q^2}{\pi^2 \Lambda_T^2}}\right)},
\label{eq:ecc}
\end{align}
where $\Lambda_T$ is the QCD scale-fixing parameter. In Ref. \cite{Jamal:2022ztl}, the variation of the Debye screening mass $m_D(T, \mu_q)$ and the strong coupling $\alpha_s(T, \mu_q)$ with chemical potential also shown in the left and right panel of Fig.\ref{fig:mu}, respectively. At a fixed temperature, the screening mass increases slightly while the strong coupling decreases as the chemical potential increases. Despite these changes, the core mathematical structure of \(\Pi^{ij}\) remains intact, meaning that only the magnitude of the modes and polarization energy loss are affected, not their intrinsic structure. As illustrated earlier in Fig.~\ref{modes_nu_aniso} from Ref.~\cite{Kumar:2017bja}, it is confirmed that the presence of finite collisions suppresses unstable modes. Conversely, Refs.~\cite{Schenke:2006xu, Romatschke:2004au, Carrington:2014bla, Jamal:2017dqs} reveal that anisotropy enhances these instabilities. Notably, in Ref.~\cite{Jamal:2022ztl}, it is demonstrated that a finite chemical potential also amplifies the instability. Fig. \ref{modes_mu_nu_aniso} from Ref.~\cite{Jamal:2022ztl} shows how unstable modes vary with a fixed collision frequency and momentum anisotropy at different chemical potentials, suggesting that unstable modes persist at higher \(k\) values in the presence of a finite chemical potential.

 \begin{figure}
		\begin{center}
	\includegraphics[height=6cm,width=6cm]{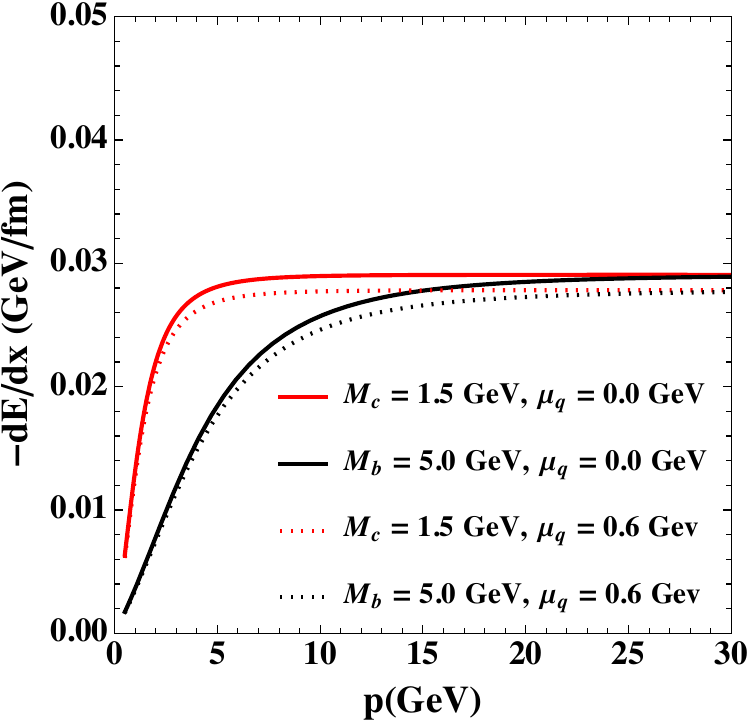} 
\caption{Comparison of polarization energy loss per unit length of charm and bottom quarks as a function of momentum, $p$, at $\mu_q = 0.0, 0.6$ GeV and $T = 0.3$ GeV ~\cite{Jamal:2020emj}.}
			\label{fig:EL_mu}
		\end{center}
	\end{figure}
Apart from Ref.~\cite{Jamal:2022ztl}, no study was found to explores collective modes in the presence of chemical potential, and the same holds true for polarization energy loss. However, Ref.~\cite{Jamal:2020emj} investigates polarization energy loss considering chemical potential and medium particle collisions. The authors in Ref.~\cite{Jamal:2020emj} employ the BGK kernel to account for particle collisions and use Eq.\eqref{eq:eq1} to introduce finite chemical potential. Their findings suggest that, at a fixed collision frequency, polarization energy loss decreases as the chemical potential increases, as depicted in Fig.~\ref{fig:EL_mu}. Moreover, no existing studies specifically address the combined effect of momentum anisotropy and medium particle collisions along with the finite chemical potential on polarization energy loss, highlighting the need for further research in this area. As discussed earlier in Sec.\ref{sec:aniso}, independent studies on anisotropic media, such as Ref.~\cite{Carrington:2015xca}, indicate that energy loss increases with anisotropy. While Refs.~\cite{Han:2017nfz, Jamal:2020fxo, Jamal:2020emj} suggest that energy loss also grows in the presence of medium particle collisions as discussed in Sec.\ref{sec:col}. This trend contrasts with the findings of Ref.~\cite{Jamal:2020emj} regarding the effect of chemical potential, where energy loss decreases with increasing chemical potential. These observations suggest a promising avenue for future research, examining the interplay between chemical potential, momentum anisotropy, and finite collision frequencies. The next section will review the possibility of energy gain in the QGP medium due to field fluctuation. 


\section{POSSIBILITY OF ENERGY GAIN DUE TO FIELD FLUCTUATION}
\label{sec:EG}
The hot QGP medium produced in HICs is a complex statistical system characterized by microscopic field fluctuations. As a heavy quark traverses this medium, it interacts with the induced field generated by its movement and encounters spontaneously generated microscopic fields, which fluctuate randomly with respect to both position and time. These fluctuations significantly impact the propagation of heavy quarks, as explored in Refs. \cite{Chakraborty:2006db, Jamal:2020fxo, Shi:2018aeb, Jamal:2021btg, Jiang:2016duz}. However, the energy loss formula employed for polarization energy loss, given in Eq. \eqref{eq:el1st} fails to capture the effects of these fluctuations, as it primarily considers only the induced fields within the plasma. To accurately reflect the influence of these random microscopic fields, it is necessary to modify Eq. \eqref{eq:el1st} and adopt a more comprehensive approach, such as the one proposed in Ref. \cite{Chakraborty:2006db} from the analogy of electromagnetic fluctuations \cite{2752149, AKHIEZER1975116} and followed in Refs.\cite{Jamal:2020fxo, Shi:2018aeb, Jamal:2021btg, Jiang:2016duz}, which accounts for both induced and fluctuating fields as:
\ba
\frac{dE}{dt}= \left \langle g_s Q^a {\bf v}(t) \cdot {\bf E}^{a}({\bf r}(t),t)\right \rangle,
\label{eq:el_fl1}
\ea
where $\left \langle .... \right \rangle$ denotes the statistical averaging. The equation of motion of the colour-charged heavy quark is written as \cite{2752149}:
\ba
\frac{{\text d{\bf v}(t)}}{{\text {dt}}}=\frac{g_sQ^a}{E_0} {\bf E}^{a}({\bf{r}}(t),t).
\label{eq:vt}
\ea

It is to note that Eq.\eqref{eq:el_fl1} can also be obtained from $\mu \ne 0$ components in Eq.\eqref{eq:1_2}. After integrating this equation, one can arrive at,

\ba
{\text {\bf v}(t)} = {\text {\bf v}_0} + \frac{g_sQ^a}{E_0} \int_{t_0}^{t}dt'{\bf E}^{a}({\bf{r}(t')},t')
\ea
\ba
{\text {\bf r}(t)} = {\text {\bf r}_0} + {\text {\bf v}_0}(t-t_0)+ \frac{g_sQ^a}{E_0} \int_{t_0}^{t}dt'\int_{t_0}^{t'}dt''{\bf E}^{a}({\bf{r}(t'')},t''),
\ea
where $E_0$, ${\bf r}_0$ and ${\bf v}_0$ are the heavy quark energy, radius-vector, and velocity at the initial
time $t_0$. For a sufficiently large time segment $\Delta t$ as compared with the period of the random fluctuations of the chromoelectric field in the QGP but small compared with the time during which the particle motion changes appreciably. Since the heavy quark trajectory differs only slightly from a straight line during this time segment, its velocity and the field acting on it at time $t = t_0 + \Delta t$ can be approximately represented by:
\ba
{\text {\bf v}(t)} \simeq {\text {\bf v}_0} + \frac{g_sQ^a}{E_0} \int_{t_0}^{t}dt'{\bf E}^{a}({\bf{r_0}(t')},t'),
\label{eq:v}
\ea
and
\ba
{\bf E}^{a}({\bf r}(t),t) & \simeq & {\bf E}^{a}({\bf r}_0(t),t) + \frac{g_sq^b}{E_0} \int_{t_0}^{t}dt'\int_{t_0}^{t'}dt''  \sum_{j}  E^{b}_j({\bf r}_0(t''),t'')\frac{\partial}{\partial r_{0j}}{\bf E}^{a}({\bf r}_0(t),t),
\label{eq:Efl}
\ea
where ${\bf r}_0(t) = {\bf v}_0 t$. Using Eq.\eqref{eq:v} and \eqref{eq:Efl} in \eqref{eq:el_fl1}, one can obtain the energy loss per unit time due to field fluctuation as:
\ba
\frac{dE}{dt}&=&\left \langle g_s~Q^a {\bf v}_0\cdot {\bf E}^a({\bf r}_0(t),t)\right \rangle + g_s^2~\frac{Q^a q^b}{E_0} \int_{t_0}^{t}dt'\left \langle {\bf E}^b({\bf r}_0(t'),t')\cdot {\bf E}^a({\bf r}_0(t),t)\right \rangle \nn
&+& g_s^2~\frac{Q^a q^b}{E_0} \int_{t_0}^{t}dt'\int_{t_0}^{t'}dt'' \bigg\langle \sum_{j}  E^{b}_j({\bf r}_0(t''),t'')\frac{\partial}{\partial r_{0j}} {\bf v_0}\cdot {\bf E}^a({\bf r}_0(t),t)  \bigg\rangle.
\label{el_full}
\ea
The above equation represents the full expression of the change in energy of heavy quarks due to induced field and field fluctuations in the QGP \cite{Jamal:2020fxo}. A detailed derivation is provided in Ref.\cite{Chakraborty:2006db}. The mean value of the fluctuating part of the field equals zero; hence $\langle {\bf E}^a({\bf r},t)\rangle$ agrees with the intensity of the field produced by the heavy quark itself in the QGP. Therefore, the first term of Eq. \eqref{el_full} corresponds to the energy change due to polarization. The other two terms correspond to the statistical change in the energy of the moving heavy quark in the medium due to the fluctuations of the chromoelectric fields and the particle's velocity under the influence of this field. The second term in Eq. \eqref{el_full} corresponds to the statistical change in the heavy quark energy associated with the correlations between the fluctuating change in the velocity of the heavy quark itself and the fluctuating chromoelectric field in the QGP. The presence of such correlations leads to a statistical magnification of the energy of the moving heavy quark. Whereas the third term corresponds to the statistical part of the dynamic friction due to the space-time correlation in the fluctuations in the chromoelectric field. Such correlations lead to additional losses (in addition to the polarization losses) in the energy of the moving heavy quark \cite{2752149, AKHIEZER1975116}.
\begin{figure}
	\includegraphics[height=6cm,width=8cm]{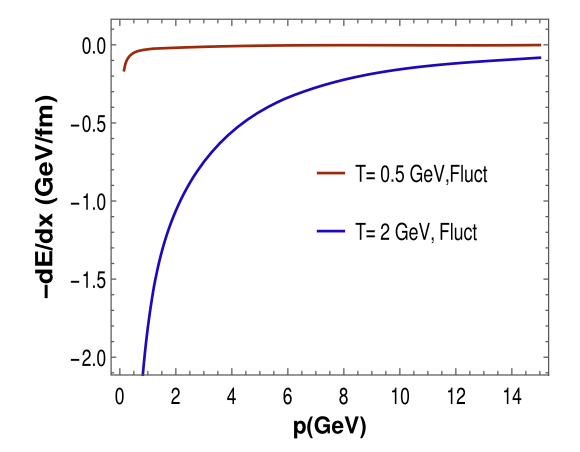}
\caption{The energy gain per unit length due to field fluctuation for the charm quark at different temperatures, T = 0.5 GeV and 2.0 GeV~\cite{Jamal:2020fxo}.}
			\label{fig_EL_FL_PRD}
	\end{figure}
\begin{figure}
	\includegraphics[height=6cm,width=6.8cm]{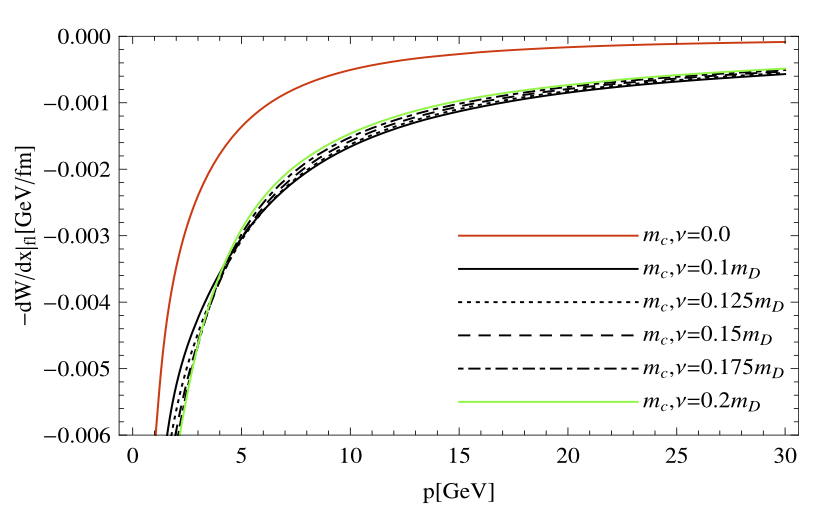}
        \includegraphics[height=6cm,width=6.8cm]{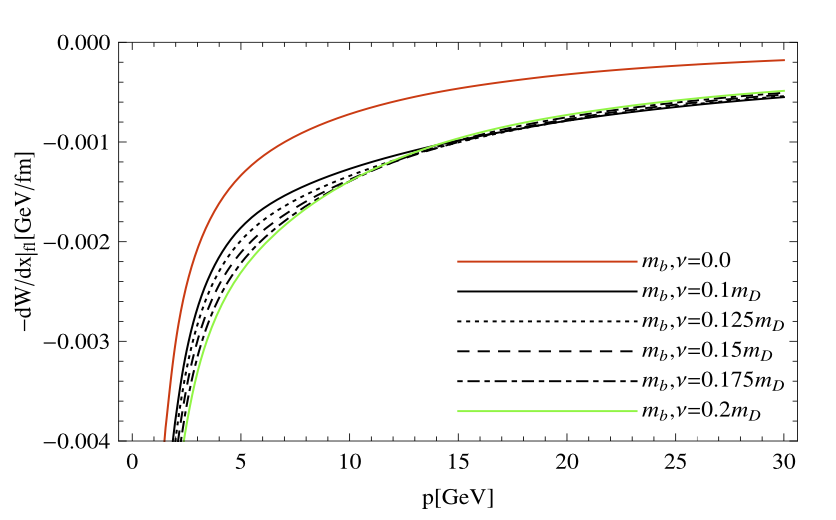}
\caption{Comparison of the energy gain per unit length, $-dW/dx = -dE/dx$ due to field fluctuation at different collision frequency $\nu = 0.0,~ 0.1~m_D$,~ $0.125~m_D$, $0.15~m_D$, ~$0.175~m_D$ and ~$0.2~m_D$ plotted as a function of momentum, $p$ for the charm quark (left panel) and the bottom quark (right panel) temperature, T = 0.3 GeV~\cite{Shi:2018aeb}.}
			\label{fig_fl_SHI}
	\end{figure}

To account for the medium effects, the chromoelectric field must be derived accordingly, which depends on the gluon propagator, as described in Eq. \eqref{eq:eind}. In Refs.\cite{Chakraborty:2006db, Jamal:2020fxo}, Eq. \eqref{el_full} is further solved by considering an isotropic QGP medium. The energy change due to field fluctuations, represented by the second and third terms of Eq. \eqref{el_full}, is evaluated in Fourier space and expressed as follows\cite{Chakraborty:2006db}:
\ba
	\frac{dE}{dt}&=&\frac{C_F \alpha _s}{\pi E_{0} |{\bf v}|^3}
	\int^{k_{\text{max}} \text{v}}_{0} \omega^2d\omega \coth\left[\frac{\beta \omega }{2}\right]F(\omega, k=\omega/v)\nn 
 &+& \frac{2C_F \alpha _s}{\pi E_{0} |{\bf v}|}
	\int^{k_{\text{max}}}_{0} dk k \int^{kv}_{0}d\omega \coth\left[\frac{\beta \omega }{2}\right]G(\omega, k),
	\label{eq:elfl}
	\ea
where 
\ba F(\omega, k) = \frac{\text {Im}[\epsilon_L(\omega,k=\omega/\text{v})]}{|\epsilon_L(\omega,k=\omega/\text{v})|^2},
\label{eq:F}
\ea
and
\ba
G(\omega,k)= \frac{\text {Im}[\epsilon_T(\omega,k)]}{|\epsilon_T(\omega,k)-k^2/\omega^2|^2}.
\label{eq:G}
\ea

In Ref. \cite{Jiang:2016duz}, medium particle collisions were incorporated using the BGK kernel, while in Ref. \cite{Jamal:2021btg}, the effects of finite chemical potential were also included. In both cases, although Eq. \eqref{eq:elfl} remained the same, the medium's dielectric permittivity, $\epsilon_L(\omega,k)$ and $\epsilon_T(\omega,k)$, are modified. Ref. \cite{Jamal:2020fxo}, as shown in Fig. \ref{fig_EL_FL_PRD}, demonstrated that the energy loss due to fluctuations is negative, signifying an energy gain. In non-equilibrium conditions, where the BGK kernel is used to account for finite collision frequencies, Ref. \cite{Shi:2018aeb} observed that heavy quarks could experience energy gain. Moreover, the medium particle collisions further enhance this energy gain compared to the no-collision case ($\nu = 0$), as illustrated in Fig. \ref{fig_fl_SHI}. Notably, this gain is prominent at low momentum but decreases as momentum rises, eventually saturating. Additionally, higher medium temperatures lead to increased energy gain. In Ref. \cite{Jamal:2021btg}, it was concluded that the presence of finite chemical potential reducing the energy gain but has a minimal impact, i.e., the gain is reduced by only about $5\%$. Importantly, in Ref. \cite{Mandal:2011bs}, the authors further explored energy loss due to field fluctuations in an anisotropic medium modeled as a two-stream system. Focusing on longitudinal chromoelectric fields for simplicity, they found that unstable modes drive an exponential increase in energy change in the presence of fluctuating fields. This suggests that field fluctuations, particularly in anisotropic plasmas, could have a significant impact on the energy dynamics of heavy quarks in HIC environments. 
\begin{figure}[ht]
	\includegraphics[height=6cm,width=8cm]{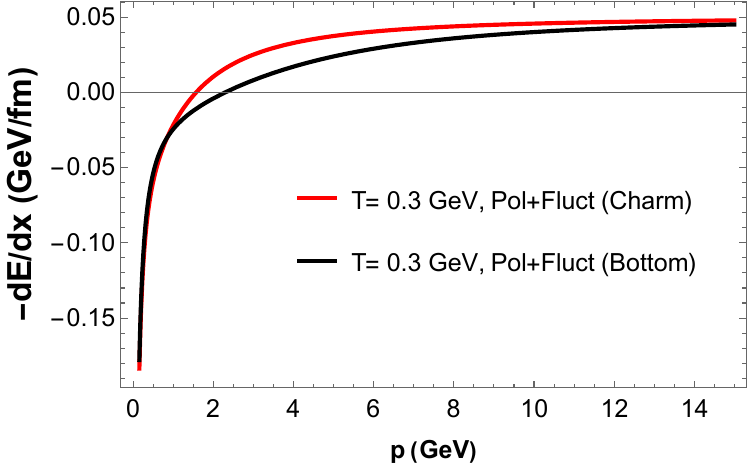} 
     \caption{Comparison of the total energy loss per unit length, $dE/dx$ of charm quark, M = 1.5 GeV and  bottom quark, M = 4.5 GeV as a function of momentum $p$~\cite{Jamal:2020fxo}.}
			\label{fig_total_EL_PRD}
	\end{figure}

The combined effects of polarization energy loss and fluctuation-induced energy gain for heavy quarks traversing the isotropic QGP were examined in Ref. \cite{Jamal:2020fxo}. The study highlighted that the energy gain from fluctuations is most significant at small momenta ($p \leq p_{\text{thermal}}$) and diminishes as the momentum increases. At higher momenta, the polarization energy loss becomes more dominant, which may explain why fluctuation energy gain has received less attention in the literature. Despite its less impact at higher momentum, fluctuation energy gain plays a crucial role in the overall energy dynamics of heavy quarks, potentially affecting key observables such as elliptic flow ($v_2$) and the nuclear modification factor ($R_{AA}$) for D and B mesons in HICs. Next, we shall discuss the incorporation of perturbative and non-perturbative effects of QGP medium through effective models and review the work done in the context of collective modes and energy loss within these models.


\section {QUASI-PARTICLE DESCRIPTION OF THE QGP MEDIUM} 
\label{sec:QP}

In this section, we will review the quasi-particle nature of the QGP medium through various models. It is important to note that not all these models are used to study collective modes and polarization loss. However, they are widely employed and provide valuable insights in different contexts. Their methodologies and implications differ slightly. We will summarize their key aspects at the end and also discuss the available literature where the quasi-particle models are used to study the collective modes and polarization energy loss.

A quasi-particle is typically conceptualized as a particle dressed by its surrounding environment, leading to modifications in its behaviour. These quasi-particles mimic actual particles in their position, momentum, and sometimes even mass. In the context of the QGP produced in high-energy HIC,  several quasi-particle models have been proposed to capture perturbative or non-perturbative medium effects to investigate various properties. These models include the effective mass model~\cite{Peshier:1995ty, Thaler:2003uz, Jeon:1995zm, Chakraborty:2010fr}, the effective fugacity model~\cite{Chandra:2007ca, Chandra:2011en}, the Polyakov loop Nambu-Jona-Lasinio (PNJL) model~\cite{Meisinger:1995ih}, the Gribov model~\cite{Gribov:1977wm, Zwanziger:1989mf}, among others. These models are instrumental in probing the intricate dynamics of the QGP and have found extensive application in examining the properties of the hot QCD medium generated in HIC. They are well documented in scientific literature. We will outline several frequently employed models and will provide a foundational understanding of their significance in modelling the QGP medium created during HIC experiments.

\subsubsection{ Effective mass Model}

The effective mass approach incorporates a temperature-dependent effective mass term to characterize the interactions of bare partons with the hot QCD medium. Inspired by concepts from solid-state physics, this approach modifies the dispersion relation for quasi-partons, expressed as \(\omega_{i} = p^{2}+m(T)_{i}^{2}\). The idea behind this approach is rooted in the assumption that effective masses, which take into account thermal interactions, permit a quasi-particle description, where these quasi-particles can either interact weakly or move freely within the medium \cite{Gorenstein:1995vm}. The resulting particle distribution functions encompass a temperature-dependent effective mass term:

\begin{equation}
f_{i}(p)= \frac{1}{e^{\beta \sqrt{p^{2}+m(T)_{i}^{2}}} \pm 1}.
\end{equation}

The effective mass approach plays a crucial role in exploring diverse properties of the hot QCD medium, encompassing both bulk and transport properties. Its versatility is evident in its application across confined and deconfined regions, underscoring its adaptability across different phases of QCD matter. Notably, the incorporation of a thermal mass term in the dispersion relation and the consequential influence of this thermal mass on the group velocity distinguish this approach from idealized scenarios. In essence, the effective mass approach offers valuable insights into the behaviour of the hot QCD medium beyond the critical temperature, underscoring the importance of temperature-dependent effective masses in comprehending its intricate properties.
Several notable studies have been conducted within the effective mass approach on the estimation of shear ($\eta$) and bulk ($\zeta$) viscosities for pure glue plasma in Ref.~\cite{Bluhm:2010qf, Chakraborty:2010fr} by utilizing the effective mass model. Ref.~\cite{Albright:2015fpa} has studied thermal conductivity, while Ref.~\cite{Puglisi:2014pda} has examined the ratio of electrical conductivity to shear viscosity. In Ref.\cite{Jeon:1995zm} authors studied the transport coefficients within effective kinetic theory.

\subsubsection{The Polyakov Loop Approach}

The Polyakov loop approach integrates gluon physics by incorporating the Polyakov loop, which is a timelike Wilson line depicted as~\cite{Polyakov:1978vu, Susskind:1979up}:

\ba 
L({\bf x}) = {\it P} \exp\left[i\int^{1/T}_{0}dx^4 A_{4}({\it t}, {\bf x})\right],
\ea
where \({\it P}\) denotes path-ordering in the imaginary time and \(A_4 = iA^0\) is the temporal gauge field, the complex Polyakov loop field \(\Phi\) is obtained by the normalized trace of \(L({\bf x})\) expressed as:

\ba
\Phi=\frac{1}{N_c}\text{Tr}_c L({\bf x}).
\ea

The thermal average of the Polyakov loop serves as an order parameter for the deconfinement transition, as established by seminal works such as those by Svetitsky and Yaffe~\cite{Svetitsky:1982gs, Svetitsky:1985ye}. Over time, various effective models have emerged to explore this concept further, with one notable extension being the Polyakov loop Nambu-Jona-Lasinio (PNJL) model, which combines the Polyakov loop with the Nambu-Jona-Lasinio model ~\cite{Nambu:1961tp, Nambu:1961fr}. The PNJL model, introduced in Ref. \cite{Meisinger:1995ih}, intertwines chiral and deconfinement order parameters, yielding predictions consistent with that of the lattice results~\cite{Tsai:2008je}. Other models, such as the Linear Sigma Model (LSM)\cite{Gell-Mann:1960mvl, Tawfik:2023dol} and its Polyakov loop extension, PLSM ~ \cite{Schaefer:2007pw}, also utilize the Polyakov loop to explore various phenomena. Moreover, the versatility of the Polyakov loop extends beyond particle physics, finding applications in cosmological studies ~\cite{Layek:2005fn, Layek:2005zu}, showcasing its applicability in diverse physical scenarios.  Refs.~\cite{Chakraborty:2010fr} have discussed a general quasi-particle theory of $\eta$ and $\zeta$ in the hadronic sector within the LSM. 

\subsubsection{The Gribov-Zwanziger approach}

The Gribov-Zwanziger model offers an alternate perspective, aiming to enhance the infrared (IR) behaviour of Yang-Mills theory by addressing residual gauge transformations overlooked by the perturbative Faddeev-Popov procedure. This model introduces a non-trivial infrared-improved dispersion relation, $\omega({\bf p})=\sqrt{{\bf p}^2 + \frac{\gamma_G^4}{{\bf p}^2}}$, incorporating the Gribov parameter, $\gamma_G$ \cite{Gribov:1977wm, Zwanziger:1989mf}. Expanding its scope, the Gribov-Zwanziger framework facilitates the evaluation of shear and bulk viscosities of the plasma, taking into account effective gluonic degrees of freedom~\cite{Florkowski:2015dmm}. Recent advancements include the development of a covariant kinetic theory formulation for the Gribov plasma, which offers insights into the derivation of transport coefficients~\cite{Jaiswal:2020qmj}. 
This approach finds application across various contexts, with notable contributions documented in references such as Refs.~\cite{Bandyopadhyay:2015wua, Canfora:2015yia, Fukushima:2013xsa, Burgio:2008jr, Dudal:2008sp}. 
Recently, Ref.~\cite{Debnath:2023zet} has adopted the Gribov-Zwanziger approach to study the soft contribution to heavy quark energy loss.

\subsubsection{Effective Fugacity Quasi-Particle Model}

The Effective Fugacity Quasi-Particle Model (EQPM) serves as a theoretical framework tailored to elucidate the characteristics of the hot QGP, particularly in the deconfined phase of QCD beyond the critical temperature, \(T_c\). In this model, effective quasi-particle distribution functions are introduced to encapsulate the medium interaction effects via effective fugacities, \(z_{g,q}\). For vanishing chemical potential, these distributions are given as~\cite{Chandra:2007ca, Chandra:2011en}:
\ba
    f_{g} &=& \frac{z_{g}\exp[-\beta E_g]}{\big(1- z_{g}\exp[-\beta E_g]\big)}, \\
    f_{q} &=& \frac{z_{q}\exp[-\beta E_q]}{\big(1+ z_{q}\exp[-\beta E_q]\big)},
\ea
where \(E_{g,q}\) denotes the energy of gluons and quarks, and \(z_{g,q}\) are temperature-dependent parameters established through non-ideal equations of state (EoSs). The quasi-particle distribution functions result in non-trivial dispersion relations, \(\omega_{i}=E_{i}+T^2\partial_T \ln(z_{i})\), where \(i\) denotes the particle species, namely gluons, quarks, or anti-quarks, effectively capturing the medium interaction effects. The EQPM finds particular relevance in the deconfined phase (\(T>T_c\)), allowing for the neglect of light quark masses.

The effective fugacities are determined by matching the pressure or energy density calculated using the EQPM with corresponding lattice or HTL results for gluonic and quark degrees of freedom individually. This procedure allows for the development of an effective transport theory, particularly in regions where weak perturbative results are trustworthy, facilitating the exploration of diverse transport and bulk properties of the QGP. The model's predictions exhibit consistency with lattice QCD results, thereby enhancing our comprehension of the QGP phase.
In Ref.~\cite{Mitra:2016zdw}, electrical conductivity and charge diffusion coefficients have been calculated using EQPM. The optical properties of the QGP medium within the EQPM approach have been studied in Refs.~\cite{Jamal:2020hpy}. The EQPM has been utilized in exploring heavy-quark transportation in isotropic medium in Ref.~\cite{Das:2012ck} and anisotropic medium in Ref.~\cite{Kurian:2019nna}. The authors in Ref.~\cite{Jamal:2018mog} discuss quarkonia dissociation within EQPM, while in Ref.~\cite{Chandra:2016dwy} they examine thermal particle production.

Additionally, there exist several models beyond those mentioned, and including all of them in this review would be impractical. Nevertheless, all of these quasi-particle models play a significant role in the process of modelling the state of QGP, thereby serving as an essential component in transport theory computations. A short comparison of the models is presented in Table~\ref{tab:1}. 

\begin {table*}
\caption {A qualitative comparison of the quasi-particle models. {Note that this comparison is based on the presently available literature. In future investigations, the applicability of the models may expand.}}
\label{tab:1}
\begin{center}
\begin{tabular}{ |p{4cm}||p{2.7cm}|p{2.7cm}|p{2.7cm}|p{2.7cm}|  }
\hline 
Properties $\downarrow$ Models $\rightarrow$ & EQPM & Polyakov Approach & Effective Mass & Gribov-Zwanziger\\
\hline 
Particle distribution function & Changes & Changes & Changes & Changes\\
\hline
Dispersion relation & Changes & Unchanged & Changes & Changes\\
\hline
Group velocity & Unchanged & Unchanged & Changes & Changes\\
\hline
Confined phase & Not Applicable & Applicable & Applicable & Applicable\\
\hline
Deconfined phase & Applicable & Applicable & Applicable & Applicable\\
\hline
Mass of medium particles & Unchanged & Unchanged & Changes & Unchanged\\
\hline
Screening mass & Changes & Changes & Changes & Changes\\
\hline
Bulk Properties & Applicable & Applicable & Applicable & Applicable\\
\hline
\end{tabular}
 \end{center}
\end{table*}

In the literature, the study of collective modes and the polarization energy loss is available within the EQPM approach. The authors in Ref.~\cite{Jamal:2017dqs} explored collective modes in the presence of momentum anisotropy considering the quasi particle scenario within EQPM. Building on this, Ref.~\cite{Kumar:2017bja} incorporated both finite collision frequency and anisotropy, utilizing EoSs derived from lattice QCD calculations~\cite{HotQCD:2014kol} and 3-loop hard-thermal-loop perturbation theory (HTLpt)~\cite{Andersen:2015eoa}. To avoid the bulk of figures, we refrain from showing them here. However, readers can refer to Refs.~\cite{Jamal:2017dqs, Kumar:2017bja} for detailed visual representations. Notably, their findings demonstrated that within the quasi-particle scenario where non-ideal medium interactions are accounted for through these EoSs—the magnitudes of all modes, including the unstable ones, are suppressed. This suggests that the quasi-particle framework effectively suppresses the instability in the QGP medium as medium particle collisions do. In contrast, the presence of finite chemical potential and momentum anisotropy  can expedite it. 

\begin{figure}[ht]
			\includegraphics[height=6cm,width=6.8cm]{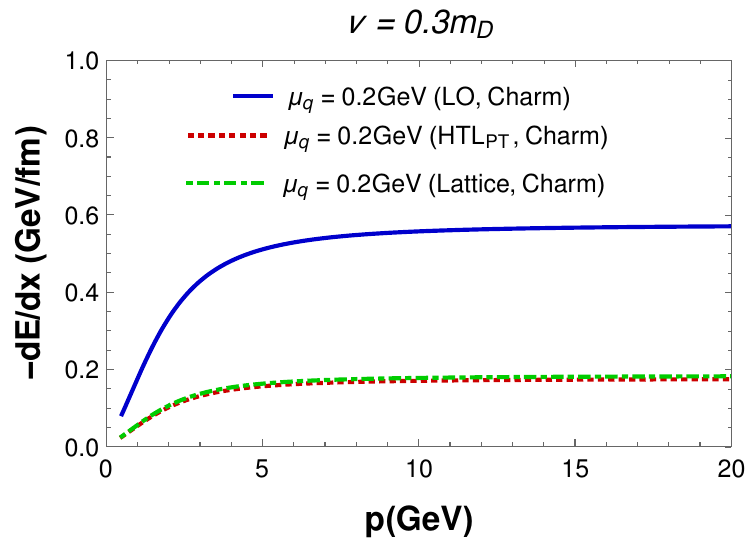} 
   \hspace{3mm}
                \includegraphics[height=6cm,width=6.8cm]{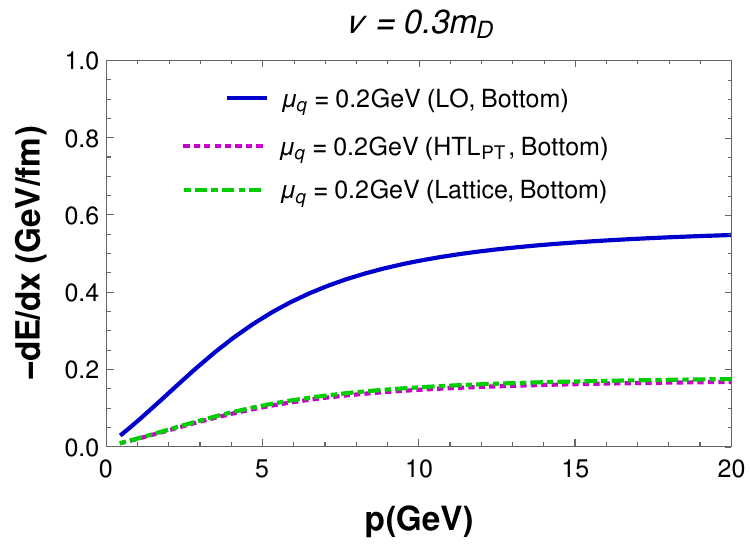} 
\caption{Comparison of the energy loss per unit length, $dE/dx$ versus momentum, $p$ for charm quark, (left panel) and bottom quark (right panel) at $\mu_q = 0.2$ GeV and $\nu =0.3 m_D$ using lattice \cite{HotQCD:2014kol} and 3-loop HTLpt \cite{Andersen:2015eoa} EoSs with LO case \cite{Jamal:2020emj}.}
			\label{fig_pol_EOS}
	\end{figure}

In Ref.~\cite{YousufJamal:2019pen}, the authors examined polarization energy loss using the EQPM while incorporating the effects of medium particle collisions. They observed that the non-ideal effects, as described through EQPM, tend to suppress the energy loss, a behaviour that contrasts with the enhancement of energy loss caused by medium particle collisions, as depicted in Fig.~\ref{fig_pol_EOS}. Furthermore, Ref.~\cite{Jamal:2020emj} extended this analysis by considering the influence of a finite chemical potential within the quasi-particle framework of EQPM. The findings revealed that, while chemical potential also suppress the energy loss, the non-ideal effects from EQPM have a more pronounced suppression. This underscores the importance of the quasi-particle approach in capturing the intricate energy loss dynamics in the QGP. These studies suggest that the quasi-particle description of QGP not only provides a more realistic picture of how the medium behaves but also emphasizes the complex interplay between different factors such as medium collisions, and chemical potential. Future research using alternative model approaches could provide deeper insights and a more comprehensive understanding of energy loss mechanisms in this medium.

\begin{figure}
			\includegraphics[height=6cm,width=6.8cm]{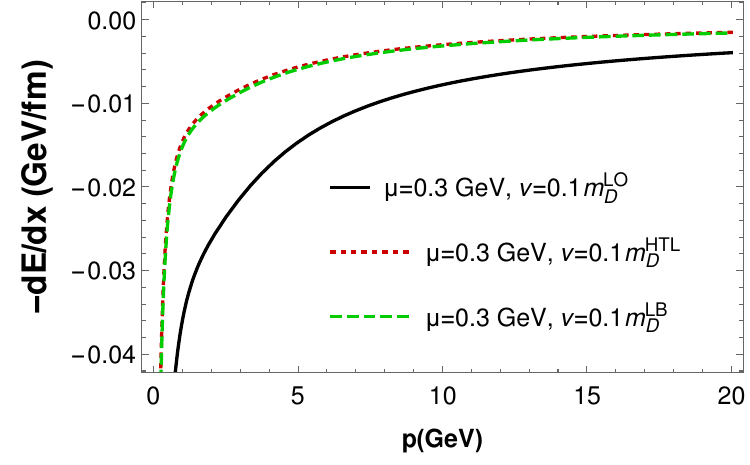} 
   \hspace{3mm}
                \includegraphics[height=6cm,width=6.8cm]{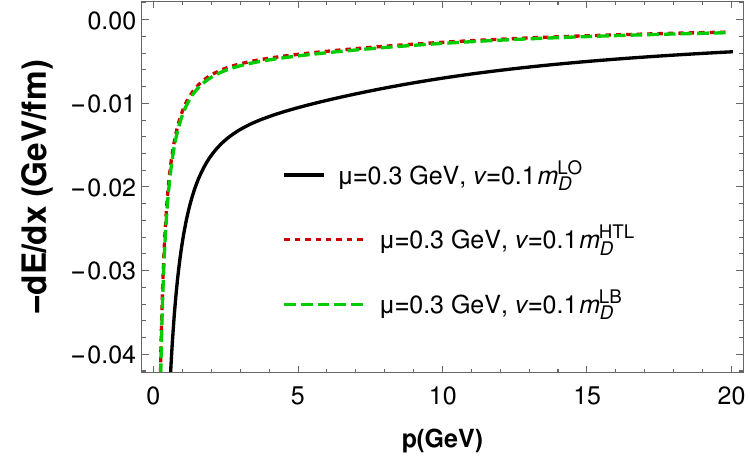}
\caption{Comparison of the energy gain per unit length, $dE/dx$ due to field fluctuation as a function of momentum, $p$ for charm quark (left panel) and bottom quark (right panel) at $\mu_q = 0.3$ GeV and $\nu =0.1 m_D$ using lattice  EoS\cite{HotQCD:2014kol} as LB and 3-loop HTLpt  EoS\cite{Andersen:2015eoa} as HTL with LO case \cite{Jamal:2021btg}.}
			\label{fig:fl_EL_BM}
	\end{figure}
 
In Fig. \ref{fig:fl_EL_BM}, taken from Ref. \cite{Jamal:2021btg}, the authors studied the energy change of moving heavy quarks due to interactions with thermal field fluctuations, considering finite collision frequency, chemical potential, and non-ideal medium effects using the EQPM. They found that, unlike in the case of finite collisions, the non-ideal medium effects suppress energy gain. In Ref. \cite{Jamal:2021btg} it is also demonstrated that the energy gain is further reduced when the presence of finite chemical potential is considered. However, the impact of non-ideal medium interactions is more significant than that of chemical potential. It would be interesting to see studies conducted within other quasi-particle models to compare these effects.

\section{Experimental Aspects}
\label{sec:EX}
This section provides a concise overview of experimental observations relevant to the ongoing discussion on the QGP medium produced in HIC experiments. Despite the formation of the QGP phase in HIC experiments, directly observing collective excitations and the energy transfer of heavy quarks at the detector end poses significant challenges. Experimental validation of heavy quark energy loss or gain is manifested in the observed hadron spectra at the detector. To understand it in very elementary words, one can assume a heavy meson ($Q{\bar q}$), i.e., consisting of a heavy quark and a light quark such as B or D -meson, formed in the initial hard collision, enters the hot QGP medium and split away. Where the heavy quark travels through the plasma, interacts with the medium, and loses/ gains energy. Upon leaving the QGP medium, it is again affixed with a light quark to form a heavy colourless meson (a coloured particle can not survive beyond $T_c$). This change in energy between the incoming and outgoing heavy meson captures through the heavy quarks energy dynamics within the QGP medium. Since, due to the presence of medium, the heavy meson comes out of the medium with a lesser energy/ momentum causes suppression of high $p_T$ heavy hadrons which subsequently affects the experimental observables such as the nuclear modification factor $R_{AA}$ and elliptic flow $v_2$ \cite{vanHees:2004gq, Lang:2012vv, rapp2010heavy, He:2022ywp, Bailhache:2024mck, Das:2024vac, Das:2022lqh, Ruggieri:2022kxv, Ghosh:2014oia, Das:2010tj}.

There are several studies where the authors have investigated heavy quark energy loss in elastic collisions, radiation, or interaction with hard modes. Various parameters such as medium anisotropy, viscosity, plasma screening, etc, are considered to comprehend these observables \cite{Fadafan:2012qu, Abbasi:2013mwa, Ghosh:2023ghi, Prakash:2023hfj, Shaikh:2021lka}. In the present case of soft contribution to the energy loss is expected to modify the ultimate output observed at the detector. Despite the limited number of studies investigating experimental observables regarding the heavy quark's soft interaction with the QGP medium, some recent investigations have shed light on this aspect. For instance, studies examining the nuclear modification factor $R_{AA}$ in this context have emerged \cite{Prakash:2023zeu, Jamal:2023ncn, Debnath:2023zet}. To calculate the $R_{AA}$, the interaction (energy loss) of heavy quarks within the QGP is quantified by drag force and momentum diffusion. These transport coefficients are employed in the Langevin equation to understand the position and  momentum evolution of the heavy quark, a process employed in several Refs.\cite{Das:2013kea, Sun:2023adv, Prakash:2021lwt, Djordjevic:2003zk, Cao:2013ita, Dutt-Mazumder:2004loa, Meistrenko:2012ju, Prakash:2024irm}:

\begin{align}
&dx_i=\frac{p_i}{E} dt,\nonumber \\
&dp_i=-\gamma_d p_i\, dt+C_{ij}\rho_j\sqrt{dt},
\end{align}
where $dt$ represents the time step, while $dp_i$ and $dx_i$ denote the changes in momentum and position of the heavy quarks, respectively. The motion of heavy quarks within the medium is influenced by two distinct forces: the dissipative/drag force, characterized by $\gamma_d$, and the stochastic force, which accounts for thermal noise ($\rho_j$). The covariance matrix $C_{ij}$ is defined as $C_{ij} = \sqrt{2D}\delta_{ij}$, where $D$ represents the diffusion coefficient of the heavy quarks. 

The authors in Refs\cite{Prakash:2023zeu, Jamal:2023ncn} determine the energy loss through the polarization process and relate it with the drag coefficient, $\gamma_d$. The relationship between energy loss and drag is given by $\gamma_d=-\frac{1}{p}\left(\ \frac{d{\text E}}{dx}\right)$, indicating that drag equals the energy loss suppressed by the momentum of the heavy quark. Subsequently, the diffusion coefficient is derived from the drag coefficients using Einstein's relation, which is expressed as $D = \gamma_d M T$ \cite{Walton:1999dy, Mazumder:2013oaa, Moore:2004tg, Debnath:2023zet}. Next, the $R_{AA}$ is defined as the ratio between the final momentum spectra, $f_{\tau_f} (p_T)$, at time ${\tau_f}$ to the initial momentum spectra, $f_{\tau_0}(p_T)$. The distribution function $f_{\tau_f}(p)$ is determined at an evolution time, $\tau_f$, while the initial spectra $f_{\tau_0}(p_T)$ is estimated within the framework of fixed-order plus next-to-leading logarithmic (FONLL) QCD corrections. This approach has been successful in reproducing the spectra of D-mesons in $p-p$ collisions following the fragmentation process, which can be parameterized as \cite{Cacciari:2005rk, Cacciari:2012ny, Prakash:2023zeu}:

\ba 
\frac{dN}{d^2p{_T}} = \frac{x_0}{(x_1+p_T)^{x_2}},
\ea
where $x_0$, $x_1$ and $x_2$ are fitting parameters. 
\begin{figure}
 \centering
\includegraphics[scale=0.32]{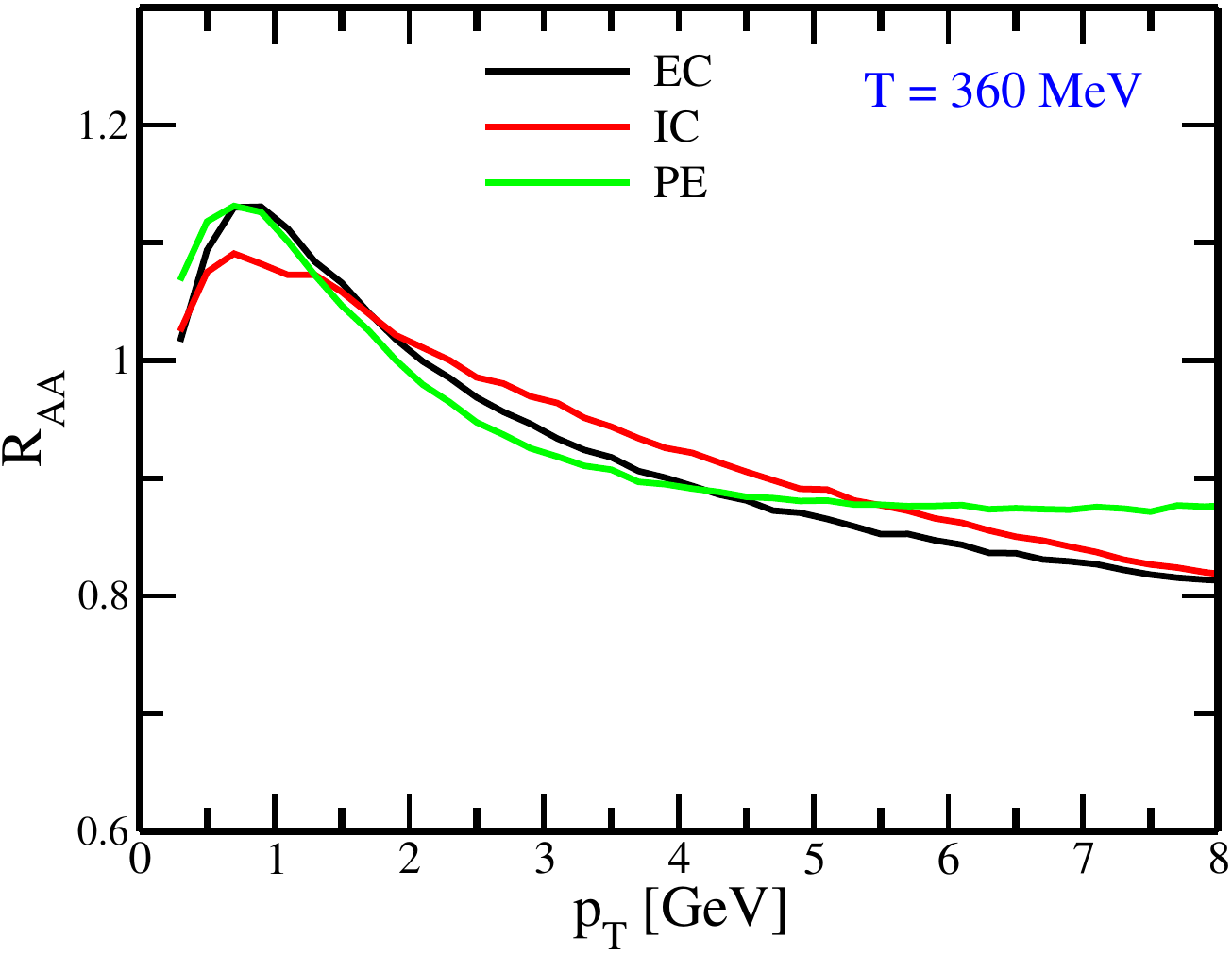}
 \hspace{1 cm}
 \includegraphics[scale=0.32]{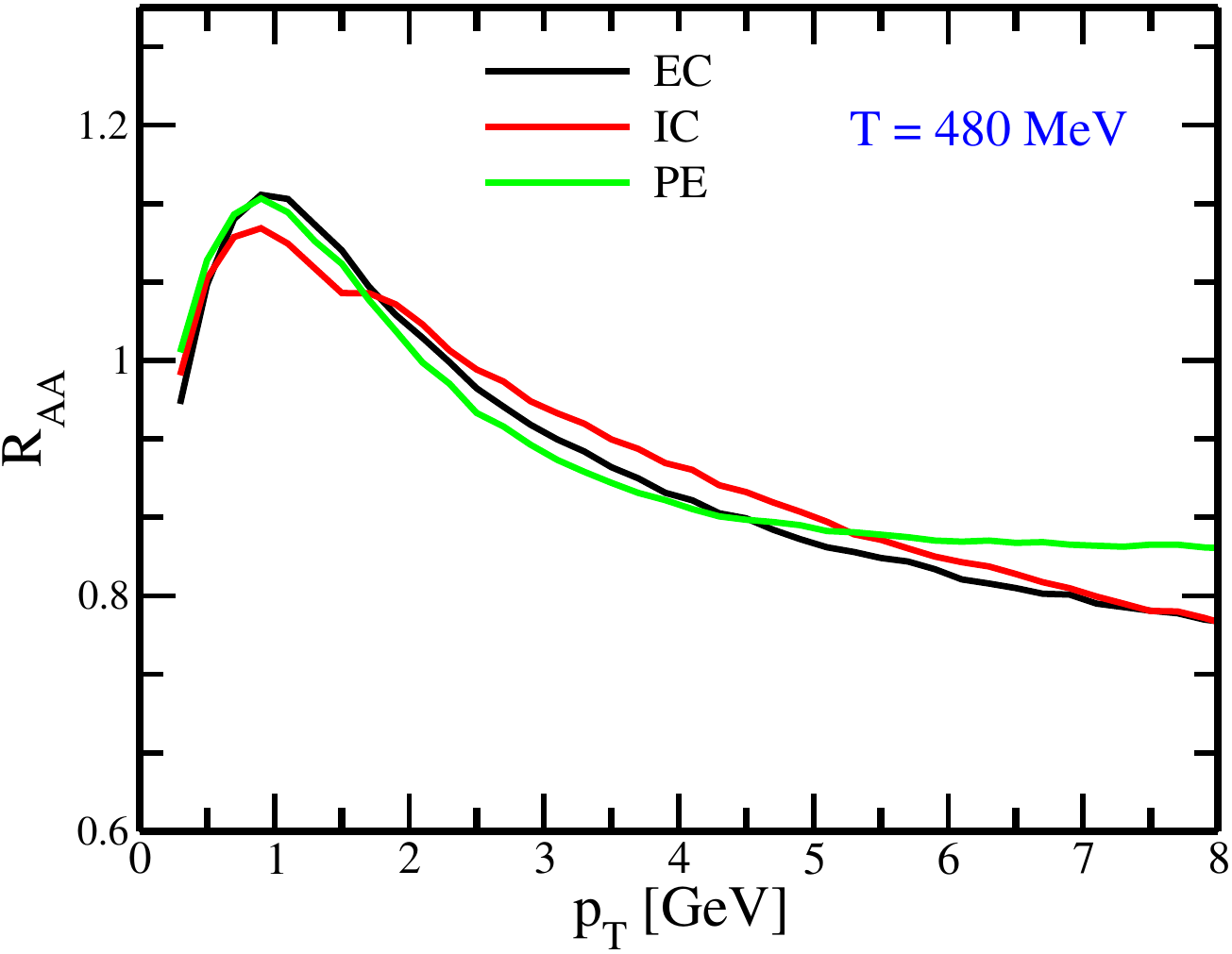}
\caption{ The $R_{AA}$ of the charm quark as a function of  $p_T$ for to PE, EC, and IC at T = 360 MeV (left panel)  and at T = 480 MeV (right panel)~\cite{Prakash:2023hfj}.}
\label{raa}
\end{figure}
\begin{figure}
 \centering
\includegraphics[scale=0.32]{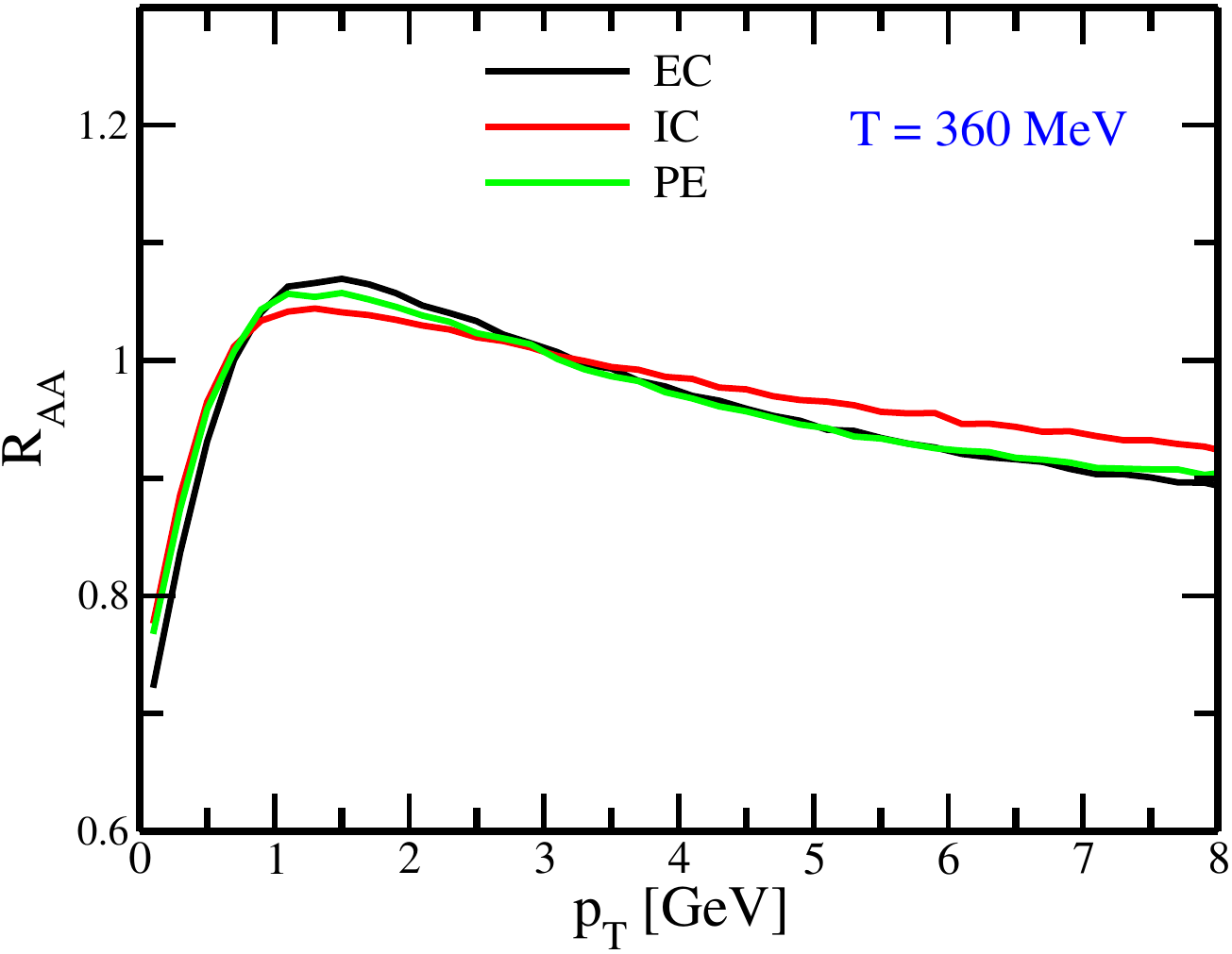}
 \hspace{1 cm}
 \includegraphics[scale=0.32]{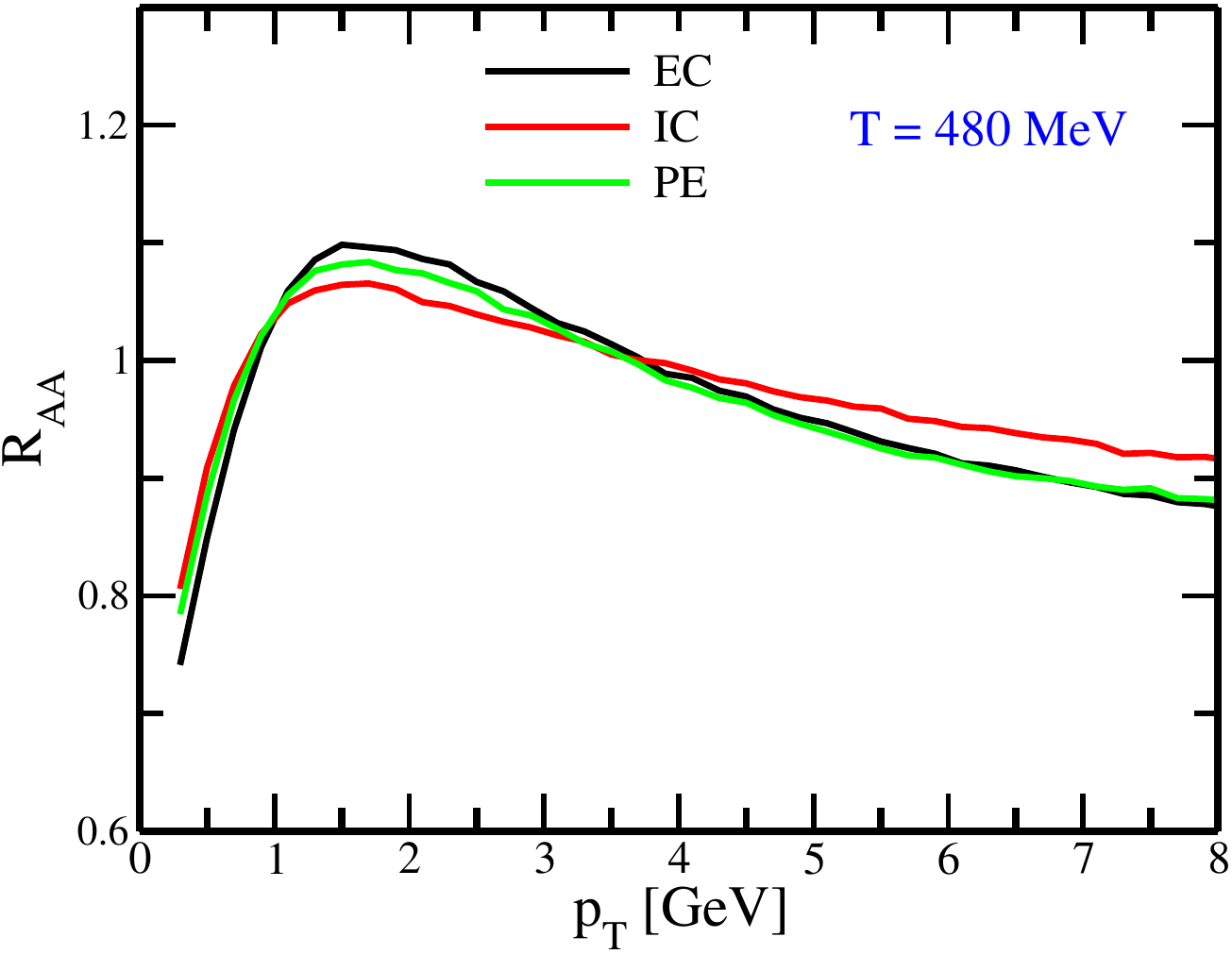}
\caption{ The $R_{AA}$ of the bottom quark as a function of  $p_T$ for to PE, EC, and IC at T = 360 MeV (left panel)  and at T = 480 MeV (right panel)~\cite{Prakash:2023hfj}.}
\label{raa_b}
\end{figure}

In Refs.\cite{Prakash:2023zeu} the authors computed and compared the $R_{AA}$ for radiation and the polarization cases and later incorporated the study of elastic collisions ~\cite{Prakash:2023hfj}, as depicted in Fig.~\ref{raa} for charm quarks and Fig.~\ref{raa_b} for bottom quarks at $T = 0.36$ GeV (left panel) and $T = 0.48$ GeV (right panel). It is claimed that at low momentum, polarization loss contributes significantly. However, at very high momentum, radiative loss dominates substantially, while the loss due to polarization saturates. Therefore, the authors concluded that it is crucial to adequately account for the contribution of polarization when studying the energy loss of heavy quarks in the medium.
However, in their analysis, the authors made the assumption of a static medium. In reality, the medium is dynamic, with observables varying over time. Thus, $R_{AA}$ and, consequently, energy loss should be determined using time-dependent parameters such as temperature to match experimental data accurately. In this context, the observation of $v_2$ is also possible, adding another layer of complexity. To date, no other studies have explored $R_{AA}$ or any other experimental aspect in the context of polarization loss. Nevertheless, considering energy gain is crucial for theoretical predictions of larger $v_2$ values, as suggested in Refs.~\cite{Rapp:2006ta, Chakraborty:2006db}.

\section{Summary}
\label{sec:Sum}
The exploration of collective excitations in QCD began over three decades ago, first introduced in Ref.~\cite{Klimov:1982bv}, and their impact on heavy quark propagation was explored approximately a decade later in Ref.~\cite{Thoma:1990fm}. Over the years, many studies have delved into these phenomena in the QGP under various conditions, including momentum anisotropy, medium particle collisions, etc. This review provides a comprehensive overview of these research efforts, focusing on how collective excitations influence the propagation of heavy quarks in the hot QGP produced during relativistic HICs. A critical aspect of these study is the role of the gluon propagator, which is essential for analyzing both the collective modes and polarization energy loss. The gluon propagator encapsulates the medium effects through its selfenergy, making it a key element in understanding these interactions. To provide clarity, this review first outlines the general framework for studying collective modes and energy loss, followed by a detailed discussion based on the available literature of these phenomena under various scenarios in different subsections.
For a complete understanding, the derivation of the gluon selfenergy in different conditions is also showed, explained its effects on the modes and energy loss. The discussion begins with the simplest case, the isotropic medium, and progressively moves to more complex scenarios involving medium particle collisions, momentum anisotropy, and their combined effects. Later sections examine the influence of finite chemical potential on these processes. In addition, the possibility for energy gain due to field fluctuations is reviewed. Notably, the quasi-particle approach have been discussed through various effective models. However, the collective modes and the energy loss/gain is only found to study within the EQPM approach. Beyond these, we also review the experimental implications of these studies. 

Certain noteworthy studies deserve special mention. For instance, investigations into the behaviour of collective excitations in a chiral medium~\cite{Ghosh:2023ghi}, in the presence of magnetic fields~\cite{Jamal:2023ncn}, and in a viscous medium~\cite{Jiang:2014oxa} offer further insights. Additionally, the energy gain due to diffusion in the evolving Glasma plays a crucial role in shaping the initial pQCD spectrum of heavy quarks before QGP formation~\cite{Ruggieri:2018rzi, Liu:2019lac}. Since the evolving Glasma represents a non-thermal equilibrium system, the diffusion of heavy colour probes~\cite{Boguslavski:2020tqz} in this medium is somewhat analogous to diffusion in Brownian motion, especially when averaged over the entire heavy quark spectrum. 
The findings discussed in this review could be qualitatively applied to the diffusion processes in the Glasma. A detailed investigation of these aspects is beyond the scope of this review and is left for future work.
 Throughout this article, several promising avenues for future research have been identified. Apart from them, most of the studies discussed here operate under the assumption of abelianization, particularly in the high-temperature limit. While this approximation has been useful for many analyses, incorporating non abelian effects could lead to more accurate and robust results. Key factors such as medium anisotropy, viscosity, vorticity, finite chemical potential, particle collisions, medium temperature, strong coupling effects, and plasma screening must be thoroughly considered in future studies to better understand these processes. Additionally, the literature on fluctuation energy gain is relatively sparse which must incorporate in the analsyis. Beyond the EQPM, further exploration of polarization energy loss and fluctuation energy gain under other quasi particle models could significantly enhance our understanding of the energy exchange mechanisms in the QGP medium.

The results reviewed in this article have important implications for experimental observables, particularly at low momentum, such as the nuclear modification factor ($R_{AA}$) at RHIC and LHC energies. These findings may also impact future experimental facilities like FAIR and NICA, which focus on A-A collisions and observables such as triggered $D\bar D$ angular correlations and heavy quark's directed flow ($v_1$)~\cite{Das:2016cwd, Nahrgang:2013saa}.  Accounting for bulk expansion of the QGP would provide a more realistic depiction of these scenarios, as suggested in Refs.~\cite{Rapp:2018qla, Dong:2019unq}. Another intriguing direction for future research is extending these analyses to the Glasma phase represents the early stages of heavy-ion collisions. However, the Glasma is a highly non-thermal system, challenging such extensions and requiring substantial modifications to the current frameworks. While some efforts have been made to explore this aspect~\cite{Jamal:2020fxo}, further investigation could yield valuable insights into the initial dynamics of heavy-ion collisions, potentially bridging the gap between early-stage evolution and QGP formation.

\section*{Acknowledgements}
MYJ acknowledges V. Chandra, S.K. Das, A. Kumar, A. Bandyopadhyay, J. Prakash and M. Kurian for the valuable discussions. MYJ acknowledges NISER for providing the postdoctoral position during the start of this work and SERB-NPDF for funding support (File No. PDF/2022/001551).  We express gratitude to all the authors of the references used in this article. We extend appreciation to the broader scientific community and the people of the country for their continued support and encouragement in fundamental scientific research.

\bibliography{ref1}   
\end{document}